\documentclass[onecolumn,usenatbib]{mn2e}

\usepackage{times}
\usepackage{graphicx}

\title[Geodesic chaos around black holes with discs or rings]
      {Free motion around black holes with discs or rings:\\
       between integrability and chaos --- I}
\author[O. Semer\'ak and P. Sukov\'a]
       {O. Semer\'ak\thanks{E-mail:
        oldrich.semerak@mff.cuni.cz} and P. Sukov\'a\\
       Institute of Theoretical Physics, Faculty of Mathematics and Physics,
       Charles University in Prague, Czech Republic}
\begin{document}

\date{}

\pagerange{\pageref{firstpage}--\pageref{lastpage}} \pubyear{}

\maketitle

\label{firstpage}

\begin{abstract}
Geodesic dynamics is regular in the fields of isolated stationary black holes. However, due to the presence of unstable periodic orbits, it easily becomes chaotic under various perturbations. Here we examine what amount of stochasticity is induced in Schwarzschild space-time by a presence of an additional source. Following astrophysical motivation, we specifically consider thin rings or discs lying symmetrically around the hole, and describe the total field in terms of exact static and axially symmetric solutions of Einstein's equations. The growth of chaos in time-like geodesic motion is illustrated on Poincar\'e sections, on time series of position or velocity and their Fourier spectra, and on time evolution of the orbital ``latitudinal action". The results are discussed in dependence on the mass and position of the ring/disc and on geodesic parameters (energy and angular momentum). In the Introduction, we also add an overview of the literature.
\end{abstract}


\section{Introduction}

It became evident no later than the 19th century that linear dynamical systems do not always represent adequately what happens in nature and society. It was also found that non-linear systems may behave in a complicated, ``chaotic" way, even if governed by fully deterministic rules and effects. And since the non-linearity occurs in so simple cases as a set of three Newtonian gravity centres, linear systems in fact proved to be just marginal tips within a vast non-linear tangle. In the 20th century, mathematical background for the dynamical systems was notably supplied by the ergodic theory and by the study of differential equations (KAM theorem). However, it was only during the past half-century that the ``theory of chaos" really established, focusing on simple non-linear dynamical systems whose orbits show highly sensitive dependence on initial conditions. Computers and numerical techniques allowed to follow and visualise long-term evolution of these systems and chaotic processes were soon detected in almost every discipline --- the days of non-linear science have come \citep{Scott-07}.

It is for about 40 years now that the chaotic behaviour has been studied in general relativity. The amount of chaos can be expected higher in such an inherently non-linear theory, allowing the space-time dynamics itself to turn irregular. The latter has naturally been examined on cosmological solutions, mainly on Bianchi IX (Mixmaster) models. While the secular debate about chaoticity and the very nature of these universes continues \citep{Coley-02,FayL-04,BeniniM-04,SoaresS-05,HeinzleRU-06,BuzziLd-07,
AndriopoulosL-08,HeinzleU-09},
one of the later attractors is to assess the effect of the cosmological constant and/vs. that of a scalar field within the FLRW-model dynamics \citep{JorasS-03,FaraoniJT-06,HrycynaS-06,LukesGBC-08,MaciejewskiPSS-08}.
(We only give several more recent references here. The early ones can be found in \cite{HobillBC-94}, for example.)

The other type of dynamical system investigated in general relativity is the test motion in a {\em given} space-time. In this field, developed over the last 20 years, several obvious lines of research arise. First, it is interesting to start from some background where the (geodesic, or electro-geodesic) flow is regular and then consider various types of its ``perturbations" in order to study whether and how much stochasticity is induced. In gravitational systems, most common ingredients are
(i) how the background field and interaction are described (in a Newtonian, pseudo-Newtonian, post-Newtonian, post-Minkowskian or exact relativistic manner; and with or without ``dark energy" contribution),
(ii) how the test body is described (mainly whether it is point-like, without structure, or endowed with more multipoles), and
(iii) whether and how the gravitational and EM radiation emissions and corresponding reactions on the body are taken into account.
Such a study actually reveals whether the respective approximations (neglects) are admissible in judgements concerning the long-term evolution of a given physical system. The ``disentanglement" of the effects of various approximations may be problematic, however, because the latter may furthermore be interlaced intricately.

To top it all, there is the ``dynamics" of the employed numerical code. With how much (and what kind of) chaos does the program itself and round-off errors confuse the results? How are these ``machine instabilities" enlaced with those coming from true physics? Interestingly, all this need not lead to desperation according to the shadowing theorem \citep{SauerGY-97} which states that, for some systems (surely for hyperbolic ones), every numerically produced ``pseudo-trajectory" is {\em shadowed} by a bunch of ``true trajectories" (all of which start from slightly different initial conditions in general) that remain close to it for a certain (``long") time (see \cite{LaiGK-99,Shi-etal-01,Judd-08} for more recent contributions). Anyway, the above uncertainties remind that it is important to have more ways of studying the dynamics and more indicators of chaos with precisely known interrelations. They also point out the preference for invariant measures of chaos and the principal role of analytical methods (like the Melnikov method, often used in relativity). Let us mention, apropos this emphasis, an important result of \cite{Motter-03}: contrary to the opinion prevailing until then, it was shown that the {\em sign} of Lyapunov exponents is coordinate independent under the conditions when the exponents can be meaningful indicators of chaos (see also the recent sequel by \cite{MotterS-09}).

One of natural backgrounds for the study of general relativistic test-particle dynamics are those corresponding to gravitational waves, because the latter have no Newtonian analogue. The waves may influence the particles in a chaotic manner, as already shown by \cite{VarvoglisP-92}; they considered charged particles moving in a uniform magnetic field and affected by a linearly polarised weak gravitational wave propagating perpendicular to the magnetic field-lines. Similar conclusion was obtained for geodesic motion in {\it pp}-waves \citep{PodolskyV-98} and recently extended to the waves described by Kundt metrics \citep{PodolskyK-07}; chaos was detected there by analysing the dimension of boundaries between the sets of initial conditions (``basins") leading to different kinds of final outcomes. (Fractal boundaries evidence chaos.)

However, it is doubly apropos to seek for chaos around the third principal ``non-Newtonian" prediction of general relativity (besides dynamical cosmology and gravitational waves) --- around black holes. Actually, in ancient Greek {\it chaos} meant a gaping bottomless void, where everything falls endlessly\dots

\section{Dynamics of test motion in black-hole fields}

Let us start a summary of previous results concerning black-hole fields by quotation from \cite{CornishF-97}:
``\dots even the most pristine black-hole space-time harbours the seeds of chaos in the form of isolated unstable orbits. A small perturbation causes these unstable orbits to break out and infect large regions of phase space. Note that the experience with Newtonian systems is very misleading. For example, the Kepler problem has more integrals of motion than are needed for integrability. Keplerian systems are thus impervious to small perturbations. In contrast, black hole space-times are at the edge of chaos, just waiting for the proverbial butterfly to flap its wings."\footnote
{The role of unstable circular orbits and their relation to homoclinic (separatrix) and zoom-whirl orbits were recently clarified, in the Kerr field, by \cite{LevinPG-09,PerezGizL-09}. The stability of circular orbits (the null ones, in particular) was also related to quasi-normal modes of black holes in the Schwarzschild(-de Sitter) and higher-dimensional (Schwarzschild-Tangherlini and Myers-Perry) space-times \citep{Cardoso-etal-09}.}

The proverbial flap has been provided by various kinds of perturbations and their effect explored by several methods.
The first perturbation we mention is on the side of the test particle; it consists in

\subsection{endowing the particle with spin,}

which deflects its motion from a geodesic and thus away from a sufficient number of conserved integrals. \cite{SuzukiM-97} performed long-term integration of the Mathisson-Papapetrou equations (describing the evolution of spinning test particles in a pole-dipole approximation) in a Schwarzschild field, in order to show, on Poincar\'e sections, that the spin really leads to chaos.
Later \cite{SuzukiM-99} analysed the imprints of particle dynamics on gravitational waves emitted and learned that in a chaotic mode the emitted power increases, the frequency spectrum is ``blurred" and the waveforms may also be changed.
\cite{KaoCh-05} confirmed the spin-induced chaos by calculating the Melnikov integral. This integral along the unperturbed homoclinic orbit from Poisson brackets of original and perturbation Hamiltonians reveals a ``transversal distance" in phase space between stable and unstable manifolds emanating from a homoclinic orbit. Infinitely repeating zeros of the integral indicate that the stable and unstable manifolds are entangled, thus evidencing the occurrence of chaos.
The same system was also studied by \cite{KoyamaKK-07} who were able to classify its chaotic motions into the ``1/frequency" and ``white-noise" types on the basis of the power spectrum of time series of the particle's $z$-position. The ``1/frequency", resp. ``white-noise" spectrum was detected for smaller, resp. larger values of the particle's spin, when the motion tends to be rather weakly, resp. strongly chaotic.

The dynamics of spinning particles was studied in the Kerr field, too.
\cite{Hartl-03a,Hartl-03b} showed, by computing the Lyapunov exponents, that it {\em may} become chaotic, but only for the spin values that are too large to be astrophysically realistic, or even too large for the pole-dipole test-particle approximation to be valid. Rotation of the hole was noticed to rather enhance the instability.
\cite{Han-08a} found the effect of the centre's rotation just opposite, by plotting Poincar\'e sections and mainly by computing a modified Lyapunov exponent --- the ``fast Lyapunov indicator".
In the meantime, \cite{KiuchiM-04} checked whether chaos reflects on emitted gravitational waves. They concluded that the spectrum of waves from non-chaotic orbits contains only discrete characteristic frequencies, whereas that from chaotic orbits is continuous with finite widths (the waveforms did not seem to be altered significantly).

Now let us turn to much wider range of options of

\subsection{perturbing the background.}
\label{perturbing-background}

Like in the above case of the test body, one can ``endow" the background (in a more or less artificial manner) with some extra multipoles. In the literature, their effects on dynamics was mostly inferred from Poincar\'e surfaces of section.
\cite{VieiraL-96a} thus showed how geodesics around a Schwarzschild black hole become chaotic under the perturbation by a quadrupole, and mainly octupole moments (in a Newtonian limit, the resulting field is an analog of the H\'enon-Heiles potential).
Chaos was also detected \citep{LetelierV-97a} around a slowly rotating black hole superposed with a dipolar halo; particles counter-rotating with respect to the black hole turned out more unstable than the co-rotating ones.
Relativistic, linearised and Newtonian core-shell fields (Schwarzschild centre plus dipole, quadrupole and octupole) were compared by \cite{VieiraL-99}. The dynamics proved to be sensitive to breaking of the reflection symmetry (disappearance of the equatorial plane) which enhanced/inhibited the chaos in oblate/prolate fields. Anyway, the relativistic dynamics was significantly more chaotic than its Newtonian counterpart.
The latter was confirmed by \cite{GueronL-01} for a black hole (vs. a point centre) superposed with a dipolar field. However, the system became even more chaotic (than the exact case) when the Schwarzschild-like centre was simulated by a pseudo-Newtonian (Paczy\'nski-Wiita) potential; this instability was amplified further by incorporating special relativistic equation of motion.
\cite{WuH-03} tested, on geodesics around the Schwarzschild hole with a dipolar shell, a new coordinate-free prescription for computation of Lyapunov exponents.
A dipolar term was also employed as a perturbation of the Reissner-Nordstr{\o}m space-time \citep{ChenW-03}; the geodesic structure again turned chaotic.
In the meantime, \cite{GueronL-02} superposed a quadrupole with a static or rotating black hole and observed that prolate deformations lead to chaos, whereas the oblate-system dynamics remained regular; in case of the rotating black hole, chaos preferentially occurred among counter-rotating orbits.
Finally we mention the usage of the basin-boundary method:
\cite{deMouraL-00a} checked the escape of free test particles from a static black hole superposed with a dipole, quadrupole and octupole terms, and found null geodesics more regular than the time-like ones. This conclusion was reached by analysing the dimension of the boundaries of initial-condition ``basins" that lead to different escape endpoints (in isolated gravitational systems, there are three possible outcomes usually: fall to the centre, escape to infinity and eternal orbiting).

Rather than to ``switch on" the multipoles individually, it is also possible to incorporate them by turning to some more general exact stationary (usually also axisymmetric) fields.
\cite{SotaSM-96} devoted their study to several {\em static} axisymmetric space-times, namely those of the Zipoy-Vorhees class describing the fields of finite axial rods (including e.g. the Schwarzschild and Curzon metrics as special cases), and also to the system of $N$ Curzon-type point singularities distributed along the symmetry axis. They diagnosed geodesic chaos both on Poincar\'e maps and Lyapunov exponents, but mainly tried to relate the known ``reliable" means of recognising chaos (detection of the homoclinic tangle, in particular) with methods based on curvature properties of the space-time background. Specifically, they studied tidal-matrix eigenvalues and curvature of the fictitious space obtained by energy-dependent conformal mapping within the Newtonian approach, while the Weyl-tensor eigenvalues in the relativistic case.
\cite{VieiraL-96b} replied, however, by demonstrating that the criterion suggested in previous paper is neither necessary nor sufficient for the occurrence of chaos. They admitted there surely exist links between global dynamics and local (curvature) properties of a certain ``configuration manifold" of the system, but concluded that ``any local analysis, in effective or even physical spaces, is far from being sufficient to predict a global phenomenon like chaotic motion". (The ``geometric" approaches have nevertheless made a considerable progress since, see concluding remarks.)
Two other, quite recent papers also illustrate differences that not so rarely occur in the field.
\cite{DubeibePS-07} analysed the geodesic dynamics of the Tomimatsu-Sato space-time and its deformed generalisation provided by a particular vacuum case of the five-parametric solution of Manko et al. (considered e.g. as a possible description of the neutron-star field). They affirmed regular dynamics in the Tomimatsu-Sato ($\delta=2$) solution, while in its generalisation they found chaos for the oblate case instead of the prolate one; this is in contrast to the results of \cite{GueronL-02}.
Soon afterwards, \cite{Han-08b} re-examined the case of the Manko et al. space-time and came to the opposite conclusion (prolate field chaotic, whereas oblate field regular).

Among the more general stationary axisymmetric black-hole backgrounds, the most important are those which might have some astrophysical relevance. These correspond to ``the most symmetric" type of perturbations --- by additional axially and reflectionally symmetric bodies (namely a disc or a ``halo").
Referring to the conditions in galactic centres, \cite{VokrouhlickyK-98} computed Newtonian trajectories around a point centre surrounded by a thin disc and monitored how the long-term evolution of their parameters responds to the disc's gravitational field as well as to successive mechanical interactions switched on to simulate crossings of the disc.
\cite{SaaV-99} solved a similar problem, with an infinite homogeneous thin disc\footnote
{Note that the potentials of such discs raise to infinity in perpendicular direction.}
around a static black hole (relativistic version) or a point mass (Newtonian version). The relativistic system turned out to be more chaotic than the Newtonian.
\cite{Saa-00} then compared the results (namely Poincar\'e sections) with the (Newtonian) case involving a thick disc (homogeneous in radial direction). Chaos were rather attenuated by the disc thickness. To the contrary, perturbations non-symmetric with respect to the disc plane (dipole) turned out to strongly enhance the chaos.
The same Newtonian system was later studied by \cite{KiuchiKM-07} on Lyapunov exponents and on time evolution of orbital parameters (and the corresponding radial-motion power spectrum). The authors mainly showed, however, that the degree of chaoticity clearly reflects on gravitational waves generated (according to the quadrupole formula) by the orbiting particles, namely on their forms and energy spectra.
We note in passing that \cite{Ramos-CaroLG-08} analysed Newtonian motion in the fields of several members of the Kalnajs disc family (sole discs, {\em without} the central body) as a possible approximation of the motion of stars in spiral galaxies. They found chaos in the case of disc-crossing orbits (just induced by discontinuity of the field, no mechanical interaction is involved), whereas regular motion otherwise.
\cite{WuZ-06} recognised, on Poincar\'e sections and Lyapunov exponents as well, geodesic chaos in the relativistic field of a Schwarzschild black hole superposed with discs of the ``peculiar" family constructed by \cite{LemosL-94} from identical halves of two dipoles.
Returning to a direct astrophysical setting: \cite{KarasS-07} proved the relevance of dynamics in the centre--disc fields for the long-term evolution of stellar clusters in galactic nuclei. The disc was found to bring the main ``instability", whereas the spherical-cluster mean field and perihelion advance (included on post-Newtonian level) rather regularise the dynamics, the total effect still being the increase of probability of close encounters between stars and the centre (probably a supermassive black hole) by a factor of 10.

Among the astrophysically important ``perturbations", there is also counting in a magnetic field.
In one of early papers, \cite{KarasV-92} discussed the dynamics of charged particles in the Ernst space-time (Schwarzschild black hole immersed in a magnetic field) with the help of Poincar\'e sections and Lyapunov exponents.
Quite recently, \cite{TakahashiK-09} plotted the Poincar\'e sections through the dynamics of charged particles in the field of a Kerr black hole surrounded by the Petterson's test dipole magnetic field. They found that the dragging effects tend to suppress the instability;\footnote
{A result that can be understood as related to this one was obtained by \cite{LetelierV-98}: in a Taub-NUT space-time, the chaos induced by dipolar perturbation is attenuated by the NUT parameter (which acts as a ``gravitomagnetic monopole", i.e. as a source of dragging).}
they explain this ``regularisation" by the fact that for fast rotating holes the integral
$\frac{1}{l}\oint\sqrt{g_{\theta\theta}u^\theta u^\theta}\,{\rm d}\tau$
of the test particle's latitudinal four-velocity over one orbital cycle (of length $l$) becomes approximately conserved.

Another important perturbation is the one by gravitational waves.
In another seminal paper, \cite{BombelliC-92} used the Melnikov method in order to detect the influence of a small periodic perturbation (corresponding to weak gravitational waves) on the geodesic structure of Schwarzschild space-time.
Later \cite{LetelierV-97b} explored, also by Melnikov method, the chaos in test motion in the Xanthopoulos solution for a Schwarzschild black hole perturbed by a gravitational wave.\footnote
{The inherent iffiness of the delicate problems of nonlinear evolution may be illustrated by putting together two of the questions raised above: (i) how chaos in the motion of particles reflects on gravitational waves emitted by them and (ii) how incident gravitational waves perturb the dynamics of particles.}

Another type of perturbation has already a long history in Newtonian celestial mechanics: perturbation of the test-particle motion in a given background by an additional distant body --- the Hill's problem.
\cite{Moeckel-92} tackled its relativistic version and showed, by calculating the Melnikov integral, that the third body perturbs the two-body homoclinic orbit into a tangle of intersecting stable and unstable manifolds, thus implying that the system went chaotic.
\cite{CornishF-97} analysed the geodesic structure of the equatorial plane of the extreme Reissner-Nordstr{\o}m black hole, perturbed by an additional distant body. They detected chaos by examining the basin boundaries and pointed out that it may e.g. lead to the increased production of gravitational waves.
Chaos was also ascertained by \cite{SelaruMCG-05}, in a Newtonian setting with the primary mass endowed with a potential simulating the Schwarzschild field.
Similarly, \cite{SteklainL-06} compared the Hill problem including the Paczy\'nski-Wiita pseudo-Newtonian (Schwarzschild-like) potential and compared it with the original Newtonian version. They applied Poincar\'e maps, Lyapunov exponents and basin boundaries on specific systems of (Sun + Earth + Moon) and (Milky Way + cluster + star) parameters and concluded that the pseudo-Newtonian systems are usually --- but not always --- more unstable than their Newtonian counterparts. Recently the authors extended the analysis to one of pseudo-Newtonian systems trying to incorporate dragging effects due to the rotation of the centre \citep{SteklainL-09}; Poincar\'e maps, Lyapunov exponents and fractal escape method revealed that the orbits counter-rotating with respect to the centre are more unstable than the co-rotating ones.

When the ``third body" is heavy and nearby, it can no longer be treated as a perturbation and a two-centre (or multi-centre) problem arises. The non-rotating centres can stay in equilibrium if both (resp. all) bear charges equal to their masses and electrostatic repulsion just balances gravitational attraction. In relativity it means black holes of the extreme Reissner-Nordstr{\o}m type; their multiple field is described by the Majumdar-Papapetrou class of exact solutions. Examination of the dynamics of free test particles in the field of two such fixed black holes by Contopoulos actually triggered the research in the whole area; a survey of the early papers was e.g. given by \cite{ContopoulosP-93}. These showed that unlike the Newtonian dynamics in the two-centre background, the relativistic problem is strongly chaotic both for photons and massive particles.
This was confirmed by \cite{DettmannFC-94} who also \citep{DettmannFC-95} discussed the appropriateness of various tests of chaos (Lyapunov exponents, topological methods, Poincar\'e sections) in general relativity on this problem.
\cite{Yurtsever-95} showed that the motion of photons in the double-centre Majumdar-Papapetrou field can be translated as geodesic motion in Riemannian space described by the spatial part of the metric. This space has negative curvature, which is known to be a source of chaos.
A similar geometric approach was tested (and compared with information from Lyapunov exponents) on static multi-black-hole space-times with cylindrical symmetry by \cite{Szydlowski-97}.
Chaotic scattering of photons by double Majumdar-Papapetrou centre was studied by \cite{Levin-99}.
A thorough discussion of photon escape from the system was provided by \cite{AlonsoRS-08}. They identified a set of unstable periodic orbits that form a fractal repeller, calculated its main characteristics (Lyapunov exponents, Hausdorff dimension and escape rate, among others) and compared them with the results of numerical simulations. They also considered the limit case when one of the black holes may be considered as a small perturbation of the other (it is light and far away).
A very thorough treatment of the subject was also presented by \cite{ContopoulosH-05}. It was mainly focused on (fractal) structure of the basins of attraction of the individual black holes and of the escape basin, and on its dependence on ratio between masses of the holes.
Several papers generalised or modified the problem.
\cite{HowardW-95} analysed Newtonian dynamics in the field of two centres with a finite dipole and discussed the nature of its chaoticity in detail.
\cite{Aguirregabiria-97} showed that in the case of {\em four} extreme centres, the scattering of charged test particles is chaotic both in Newtonian and in relativistic description.
\cite{CornishG-97} also took into account a dilaton field and observed transition between regular and chaotic motion for photons and for extremally charged particles. (They added a useful discussion on fractal, topological and curvature methods.)
\cite{deMouraL-00b} considered a slightly different (and more realistic) system --- of two {\em uncharged} black holes, but far away from each other. The approximated motion of photons in this field shows chaotic features: the boundary between capture, orbital and escape basins is fractal, becoming a self-similar Cantor set in the limit of infinite distance between the centres, with its box-counting dimension falling off logarithmically.
\cite{HananR-07} recently generalised the problem of multiple extreme black holes to higher-dimensional space-times (with special attention to the 5-dimensional case) and found qualitative similarity of chaos in test dynamics there.

Finally, let us quote the variant of the orbital dynamics where the own field of the orbiting body is {\em not} negligible. The problem thus ceases to count among those with {\em given} background and involves dynamics of the Einstein equations themselves. Compact-binary evolution is a prominent astrophysical example of such a challenge. It in fact embodies, in a true {\em tangle}, all the knottiness both on the general-relativity side and on the side of non-linear dynamics.
The chaotic evolution of compact binaries was discussed, on second post-Newtonian level including spins, by \cite{Levin-00,Levin-03} and by \cite{WuX-08}; in the last paper, the ``fast Lyapunov indicator" was computed in order to deduce the roles of system parameters. (A great deal of information is also given in the 3rd PN treatment by \cite{GrossmanL-09}.)
\cite{CornishL-03} also compared it (in the case with zero spins) with {\em test}-particle dynamics in a Schwarzschild background. For the chaotic, post-Newtonian system the Lyapunov time was sometimes found comparable to the timescale for dissipation due to gravitational waves, which would indicate possible abating of chaos due to waves (were the Lyapunov time longer, chaos would be damped).
However, the conclusiveness of any post-Newtonian approximation in the issue of chaos (of course) remains a question \citep{Levin-06}.

In the present paper we study the dynamics of time-like geodesics in exact space-times describing the fields of static and axially symmetric black holes surrounded by thin discs or rings. We try to show, on Poincar\'e sections, on time series of position or velocity and their Fourier spectra, and on behaviour of the ``latitudinal-action", how the original, completely integrable Schwarzschild geodesic dynamics grow chaotic when the external source is being ``turned on", and we check how the evolution of phase-space properties depend on the values of geodesic constants.

\section{Static black hole with a ring or disc}

We are interested in time-like geodesic dynamics in the field of a black hole surrounded by a ring or a thin disc. Let us restrict to the static and axially symmetric case and to discs without charges and currents and without radial pressure. For such sources, the space-time metric can be put into the Weyl form\footnote
{Geometrised units are used in which $c=G=1$. Cosmological constant is set to zero. Index-posed commas mean partial differentiation.}
\begin{equation}  \label{Weylmetric}
  {\rm d}s^2=-e^{2\nu}{\rm d}t^2
             +\rho^2 e^{-2\nu}{\rm d}\phi^2
             +e^{2\lambda-2\nu}({\rm d}\rho^2+{\rm d}z^2)
\end{equation}
in (Weyl) cylindrical-type coordinates $(t,\rho,z,\phi)$, with the two unknown functions $\nu$, $\lambda$ only depending on $\rho$ and $z$. The ``gravitational potential" $\nu$ satisfies the Laplace equation, so it superposes linearly, while $\lambda$ is found by quadrature
\begin{equation}  \label{lambda}
  \lambda=
  \int_{\rm axis}^{\rho,z}
      \rho\left\{\left[(\nu_{,\rho})^2-(\nu_{,z})^2\right]{\rm d}\rho
                 +2\nu_{,\rho}\nu_{,z}{\rm d}z\right\},
\end{equation}
computed along any line within the vacuum region. (The requirement that the space is locally Euclidean on the axis implies $\lambda=0$ on its part lying outside the horizon.)

\subsection{Schwarzschild black hole}

The Schwarzschild field (of mass $M$) is described by
\begin{eqnarray}
  \nu&=&\frac{1}{2}\,\ln\frac{d_1+d_2-2M}{d_1+d_2+2M}
      = \frac{1}{2}\,\ln\left(1-\frac{2M}{r}\right),\\
  \lambda&=&\frac{1}{2}\,\ln\frac{(d_1+d_2)^2-4M^2}{4\Sigma}
          = \frac{1}{2}\,\ln\frac{r(r-2M)}{\Sigma}\;,
  \label{lambdaSchw}
\end{eqnarray}
where
\begin{eqnarray*}
  d_{1,2}&\equiv&\sqrt{\rho^2+(z\mp M)^2}=r-M\mp M\cos\theta,\\
  \Sigma\equiv d_1 d_2
        &=&[(\rho^2+z^2+M^2)^2-4z^2 M^2]^{1/2}
         = (r-M)^2-M^2\cos^2\theta.
\end{eqnarray*}
The second expressions are in Schwarzschild coordinates $(r,\theta)$ which are related to the Weyl coordinates by
\[\rho=\sqrt{r(r-2M)}\,\sin\theta, \;\;\;\;
  z=(r-M)\,\cos\theta,\]
or, in the opposite direction,
\[r-M=\frac{1}{2}(d_2+d_1), \;\;\;\;
  M\cos\theta=\frac{1}{2}(d_2-d_1).\]

Superpositions of the Schwarzschild black hole with an additional static and axially symmetric source are often being represented in Schwarzschild coordinates. Outside of the thin source with no radial pressure, the complete metric reads
\begin{equation}  \label{complete-metric}
  {\rm d}s^2=
    -\left(1-\frac{2M}{r}\right)e^{2\hat\nu}{\rm d}t^2
    +\frac{e^{2\hat\lambda-2\hat\nu}}{1-\frac{2M}{r}}\;{\rm d}r^2
    +r^2 e^{-2\hat\nu}
     (e^{2\hat\lambda}{\rm d}\theta^2+\sin^2\theta\,{\rm d}\phi^2),
\end{equation}
where
$\hat\nu(r,\theta)$ is the potential of the external source and
$\hat\lambda(r,\theta)\equiv\lambda-\lambda_{\rm Schw}$ ($\lambda_{\rm Schw}$ is given by (\ref{lambdaSchw})).
In the present paper, the Bach-Weyl ring and several thin discs described by exact solutions of Einstein equations will be chosen as sources surrounding the Schwarzschild black hole in a reflectionally symmetric manner (they will lie in the ``equatorial plane" of the hole).

\subsection{Bach-Weyl ring}

The thin-ring solution of \cite{BachW-22} is given by complete 1st-kind Legendre elliptic integral
$K(k)\equiv\int_0^{\pi/2}\frac{{\rm d}\phi}{\sqrt{1-k^2\sin^2\phi}}\,$:
\begin{eqnarray}
  \nu_{\rm BW}&=&-\,\frac{2{\cal M}K(k)}{\pi l_2}\,, \label{nu,Bach} \\
  \lambda_{\rm BW}&=&\frac{{\cal M}^2 k^4}{4\pi^2 b^2\rho}\,
                     \left[(\rho+b)(-K^2+4k'^2 K\dot{K}+4k^2 k'^2\dot{K}^2)
                           -4\rho k^2 k'^2(k'^2+2)\,\dot{K}^2\right],
\end{eqnarray}
where ${\cal M}$ is the ring's mass and $b$ its Weyl radius,
$\dot{K}\equiv\frac{{\rm d}K}{{\rm d}(k^2)}\,$,
$k'^2=\frac{(l_1)^2}{(l_2)^2}\,$,
$k^2=1-k'^2=\frac{4\rho b}{(l_2)^2}\,$, and
$l_{1,2}\equiv\sqrt{(\rho\mp b)^2+z^2}\,$.

\subsection{Inverted counter-rotating Morgan-Morgan discs}

A physically reasonable family of annular thin-disc solutions was obtained by \cite{LemosL-94} by inversion (with respect to the rim) of the finite Morgan-Morgan discs interpreted in terms of two non-interacting streams of dust counter-orbiting on circular geodesics \citep{Semerak-03}. The Newtonian densities of inverted discs are ($m=1,2,\dots$)
\begin{equation}
  w^{(m)}(\rho)=
   \frac{2^{2m}(m!)^{2}{\cal M}b}{(2m)!\;\pi^{2}\rho^{3}}
   \left(1-\frac{b^{2}}{\rho^{2}}\right)^{\!m-1/2},
\end{equation}
where ${\cal M}$ is mass and $b$ is the Weyl inner radius. The corresponding potential is quite lucid in oblate spheroidal coordinates $(x,y)$ which are related to the Weyl coordinates by
\begin{equation}  \label{oblate}
  \rho^{2}=b^{2}(x^{2}+1)(1-y^{2}), \;\;\;\;
  z=bxy:
\end{equation}
\begin{equation}  \label{nu,iMM}
  \nu^{(m)}(x,y)=
 -\frac{2^{2m+1}(m!)^{2}{\cal M}}{\pi\,b} \;
  \frac{\sum_{n=0}^{m}
        C^{(m)}_{2n}\;
        {\rm i}Q_{2n}\!\!\left(\frac{{\rm i}|y|}{\sqrt{x^{2}+1-y^{2}}}\right)
               P_{2n}\!\!\left(\frac{x}{\sqrt{x^{2}+1-y^{2}}}\right)}
        {\sqrt{x^{2}+1-y^{2}}} \;\; ,
\end{equation}
where (for $n\leq m$)
\[C^{(m)}_{2n}\equiv\frac{(-1)^{n}(4n+1)(2n)!(m+n)!}
                         {(n!)^{2}(m-n)!(2m+2n+1)!}\]
and $P_{2n}$, $Q_{2n}$ denote Legendre polynomials and Legendre functions of the second kind.
Note that in computations it is convenient to express the latter as
\[Q_{2n}({\rm i}X)=
  -{\rm i}\,P_{2n}({\rm i}X)\,{\rm arccot}\/X
  -\sum_{j=1}^{2n}\frac{1}{j}\,P_{j-1}({\rm i}X)P_{2n-j}({\rm i}X).\]

\subsection{Annular discs with power-law density profile}

Another family of annular thin discs is the one with power-law radial profile of density \citep{Semerak-04},
\begin{equation}
  w^{(m,n)}(\rho)=\frac{\left(1+\frac{1}{n}\right)_{m}}{m!}
                  \frac{{\cal M}b}{2\pi\rho^3}
                  \left(1-\frac{b^n}{\rho^n}\right)^{\!m},
\end{equation}
where $m$ and $n$ are natural numbers and $(a)_{m}\equiv\Gamma(a+m)/\Gamma(a)$ is the Pochhammer symbol. The corresponding fields behave somewhat more regular at the inner rim than those of the inverted Morgan-Morgan discs. For even $n$,\footnote
{When $n$ is odd, the potentials comprise logarithmic terms. We will not consider such a case here.}
the potential can be expanded as
\begin{equation}  \label{nu(rho,z),z<b}
  \nu^{(m,n)}(\rho,z)=
  -\frac{\left(1+\frac{1}{n}\right)_{m}}{n}\frac{{\cal M}}{b}
   \sum\limits_{j=0}^{\infty}
   \frac{P_{2j}(0)P_{2j}\!\!\left(\!\frac{z}{\sqrt{\rho^2+z^2}}\!\right)
         (\rho^2+z^2)^{j}}
        {\left(\frac{2+2j}{n}\right)_{\!m+1}b^{2j}}
\end{equation}
in the ``inner region" ($\sqrt{\rho^2+z^2}<b$), while as
\begin{equation}  \label{nu(rho,z),z>b,even-n}
  \nu^{(m,n)}(\rho,z)=
 -\frac{\left(1+\frac{1}{n}\right)_{m}{\cal M}}{\sqrt{\rho^2+z^2}}
  \left[
  \sum\limits_{j=0}^{\infty}
  \frac{P_{2j}(0)P_{2j}\!\!\left(\!\frac{z}{\sqrt{\rho^2+z^2}}\!\right)
        b^{2j}}
       {n\left(\frac{1-2j}{n}\right)_{\!m+1}(\rho^2+z^2)^{j}}+
  \sum_{k=0}^{m}
  \frac{(-1)^k
        Q_{kn+1}(0)
        P_{kn+1}\!\!\left(\!\frac{|z|}{\sqrt{\rho^2+z^2}}\!\right)
        b^{kn+1}}
       {k!(m-k)!\,(\rho^2+z^2)^{\frac{kn+1}{2}}}\right]
\end{equation}
in the ``outer region" ($\sqrt{\rho^2+z^2}>b$).

\subsection{Complete fields}

\begin{figure}
\includegraphics[width=\textwidth]{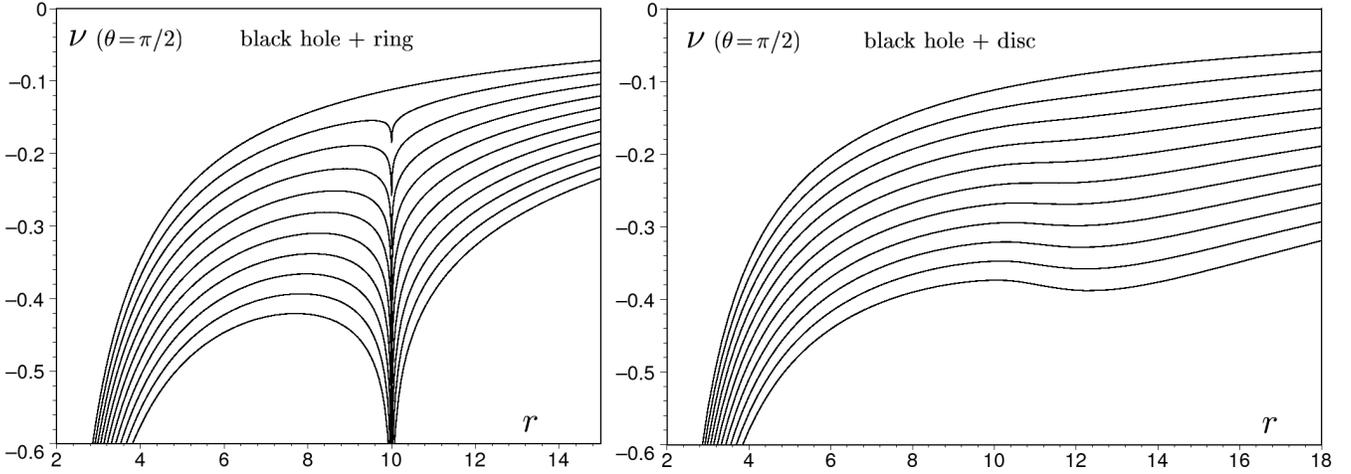}
\caption
{Radial dependence of the equatorial potential of a Schwarzschild black hole surrounded by a Bach-Weyl ring (\ref{nu,Bach}) on Schwarzschild radius $r=10M$ ({\it left}), resp. by a power-law disc (\ref{nu(rho,z),z<b},\ref{nu(rho,z),z>b,even-n}) (of $m=4$, $n=10$) with Schwarzschild inner radius $r=10M$ ({\it right}).
The relative masses of the ring are ${\cal M}/M=0$, $0.2$, $0.4$, \dots, $2.0$ (the singular ``valley" of course becomes more profound along this sequence), while the relative masses of the disc are ${\cal M}/M=0$, $0.5$, $1.0$, \dots, $5.0$. It is seen that the disc is only scarcely able to develop its own potential minimum near the black hole, even though it is fairly concentrated (towards the rim) and its mass is chosen extremely large. Radius is in the units of $M$, potential is dimensionless.}
\label{BH-ringdisc-nu-equat}
\end{figure}

The complete fields are given by sum of the potentials,
$\nu=\nu_{\rm Schw}+\hat{\nu}$. The Schwarzschild singularity is the strongest possible source of gravitation, so it is not surprising that the additional sources can hardly affect its field considerably without being unrealistically heavy (and compact). Figure \ref{BH-ringdisc-nu-equat} illustrates this on the Bach-Weyl ring (one-dimensional source) and on one of the (rather concentrated) power-law discs (two-dimensional source). Extended (three-dimensional) sources would have even weaker immediate influence on the black-hole field.

\section{Time-like geodesics in Weyl fields}

\begin{figure}
\includegraphics[width=\textwidth]{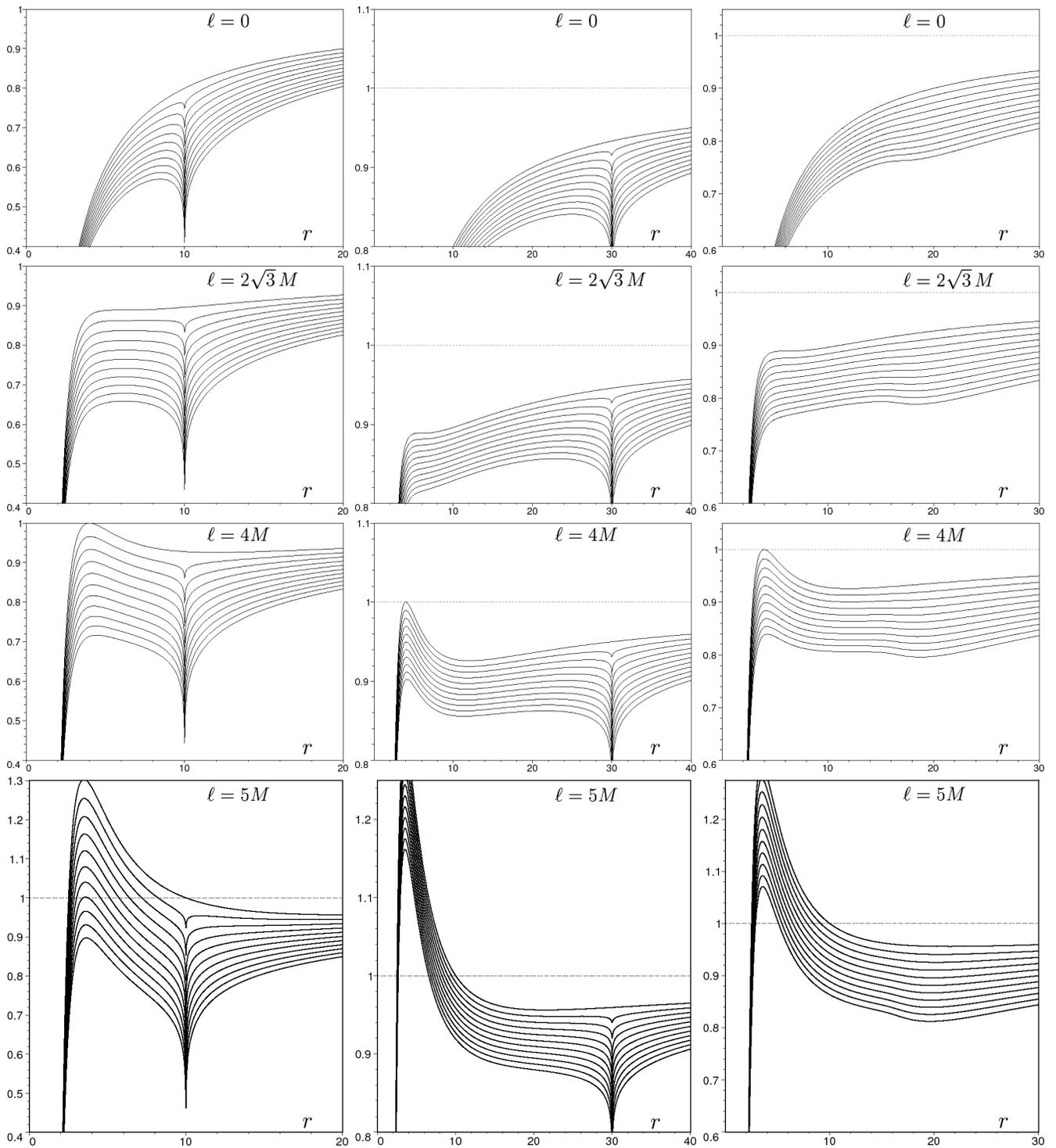}
\caption
{Radial shape of the effective potential squared (\ref{Veff}) in the equatorial plane of a Schwarzschild black hole surrounded by a Bach-Weyl ring on Schwarzschild radius $r=10M$ ({\it left column}) and $r=30M$ ({\it middle column}) and by a power-law disc (of $m=4$, $n=10$) with Schwarzschild inner radius $r=15M$ ({\it right column}). Relative masses of the ring are ${\cal M}/M=0$, $0.1$, $0.2$, \dots, $1.0$, while relative masses of the disc are ${\cal M}/M=0$, $0.2$, $0.4$, \dots, $2.0$ (the curves shift downwards with increasing mass of the additional source; the topmost curves correspond to a pure Schwarzschild black hole). From top to bottom row, the specific angular momentum $\ell$ is set to $0$, $2\sqrt{3}M$, $4M$ and $5M$, where $M$ is the black-hole mass.
Radius is in the units of $M$, the effective potential is dimensionless (per particle's rest mass). Note that display proportions are only kept fixed within the columns.
Basic observations:
(i) depending on parameters of the external source, several different configurations of circular orbits are possible; both types of these (unstable and stable) may lie either above or below the source;
(ii) the ring being much ``stronger" (singular) source than the disc, it {\em always} induces an unstable circular orbit below;
(iii) there may exist {\em two} different unstable circular orbits (and a stable one in between) between the horizon and the ring/disc; a marginal case also exists when the stable and the ``outer" unstable of these orbits coalesce into a marginally stable one.}
\label{BH-ringdisc-Veff-equat}
\end{figure}

\begin{figure}
\includegraphics[width=\textwidth]{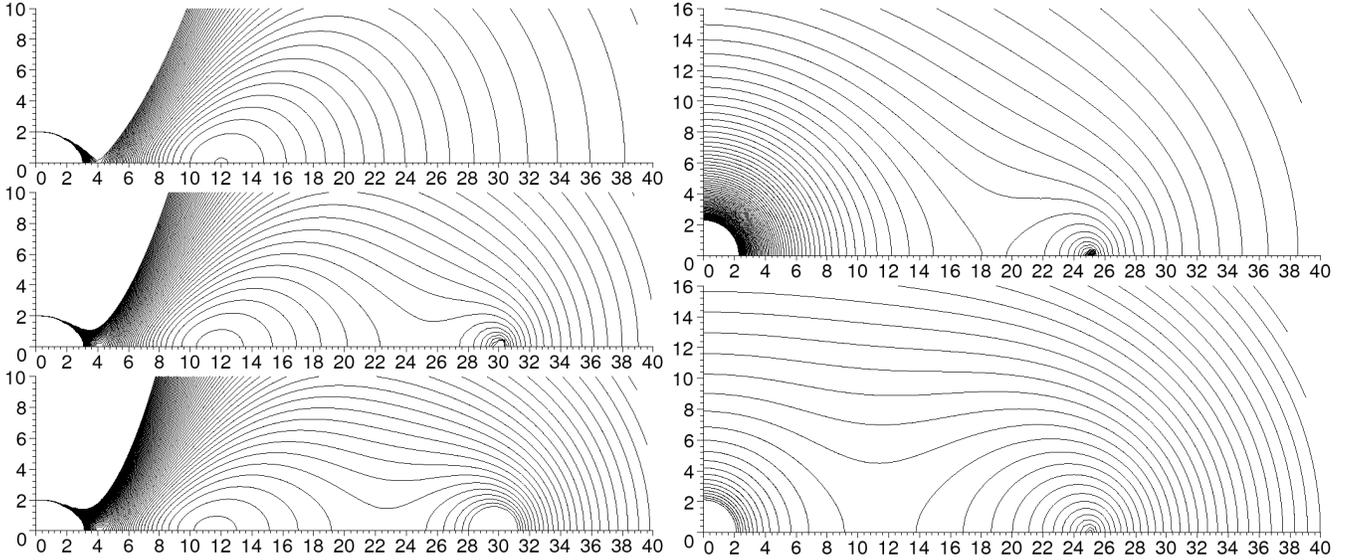}
\caption
{Contour lines of the effective potential squared (\ref{Veff}) in the meridional plane ($r\sin\theta$, $r\cos\theta$) of a Schwarzschild black hole surrounded by a Bach-Weyl ring. In the {\it left column}, the ring is on Schwarzschild radius $r=30M$ and has mass ${\cal M}=0$, $0.4M$, $0.8M$ (from top to bottom); the angular momentum is $\ell=4M$.
In the {\it right column}, the ring is on Schwarzschild radius $r=25M$ and has mass ${\cal M}=2M$, $20M$; the angular momentum is $\ell=0$.
(We have chosen such unrealistic values of the ring mass in order to show that even the singular ring has to be {\em extremely heavy} in order to attract locally as strongly as the hole.)
The coordinates are in the units of the black-hole mass $M$.}
\label{BH-ring-Veff-merid}
\end{figure}

Massive free test particles move along time-like geodesics. In Weyl fields these read, in the Weyl coordinates,
\begin{eqnarray}
  \frac{{\rm d}u^t}{{\rm d}\tau}
  &=&\frac{{\rm d}}{{\rm d}\tau}\!\left(\frac{{\cal E}}{-g_{tt}}\right)
   =-\frac{2{\cal E}}{e^{2\nu}}\,(\nu_{,\rho}u^\rho+\nu_{,z}u^z),
     \label{ddt} \\
  \frac{{\rm d}u^\phi}{{\rm d}\tau}
  &=&\frac{{\rm d}}{{\rm d}\tau}\!\left(\frac{\ell}{g_{\phi\phi}}\right)
   = \frac{2e^{2\nu}\ell}{\rho^3}
              \left[\rho(\nu_{,\rho}u^\rho+\nu_{,z}u^z)
                    -u^\rho\right],
     \label{ddphi} \\
  \frac{{\rm d}u^\rho}{{\rm d}\tau}
  &=&-\frac{{\cal E}^2\nu_{,\rho}}{e^{2\lambda}}
     +\frac{\ell^2 e^{4\nu}}{\rho^3 e^{2\lambda}}\,(1-\rho\nu_{,\rho})
     +(\nu_{,\rho}-\lambda_{,\rho})\left[(u^\rho)^2-(u^z)^2\right]
     +2(\nu_{,z}-\lambda_{,z})u^\rho u^z,
     \label{ddrho} \\
  \frac{{\rm d}u^z}{{\rm d}\tau}
  &=&-\frac{{\cal E}^2\nu_{,z}}{e^{2\lambda}}
     -\frac{\ell^2 e^{4\nu}\nu_{,z}}{\rho^2 e^{2\lambda}}\,
     -(\nu_{,z}-\lambda_{,z})\left[(u^\rho)^2-(u^z)^2\right]
     +2(\nu_{,\rho}-\lambda_{,\rho})u^\rho u^z,
     \label{ddz}
\end{eqnarray}
where $g_{\mu\nu}$ is the metric tensor,
$u^\mu\equiv\frac{{\rm d}x^\mu}{{\rm d}\tau}$ is the test-particle four-velocity, $\tau$ is its proper time and
\[{\cal E}\equiv E/m\equiv -u_t=e^{2\nu}u^t, \;\;\;\;
  \ell\equiv L/m\equiv u_\phi=\rho^2 e^{-2\nu}u^\phi\]
are constants of the motion (energy and azimuthal angular momentum per unit mass with respect to an observer staying at rest at spatial infinity).
In the vacuum regions (which means everywhere when there are no extended sources), the derivatives of $\lambda$ can be expressed as
\[\lambda_{,\rho}=\rho(\nu_{,\rho})^2-\rho(\nu_{,z})^2, \;\;\;\;
  \lambda_{,z}=2\rho\nu_{,\rho}\nu_{,z}\]
from the Einstein field equations. (However, $\lambda$ itself must in general be found numerically at each point by integrating this gradient.)
Both the metric functions are continuous everywhere outside the horizon, they only have a finite jump in normal derivatives on the disc which is given by the ``Newtonian density" of the disc $w(\rho)$ (the one standing in the Poisson equation),
\[\nu_{,z}(\rho,z\rightarrow 0^+)=-\nu_{,z}(\rho,z\rightarrow 0^-)=
  2\pi w(\rho).\]
Note that the equations for $u^t$ and $u^\phi$ do not contain $\lambda$, yet the ($t,\phi$)-motion is not fully determined by $\nu$, being coupled to the meridional components $u^\rho$, $u^z$ which depend on $\lambda$ explicitly.

The additional-source effect on geodesic motion can be tentatively estimated on effective potential for radial and/or latitudinal motion. From the four-velocity normalisation and using the metric (\ref{complete-metric}), one has
\begin{equation}  \label{Veff}
  e^{2\hat{\lambda}}\left[(u^r)^2+r(r-2M)(u^\theta)^2\right]=
  {\cal E}^2-({\cal V}_{\rm eff})^2,
  \;\;\;\;\;\; {\rm where} \;\;\;\;
  ({\cal V}_{\rm eff})^2\equiv
  \left(1-\frac{2M}{r}\right)
  \left(1+\frac{\ell^2 e^{2\hat{\nu}}}{r^2\sin^2\theta}\right)
  e^{2\hat{\nu}} \;.
\end{equation}
Above the black-hole horizon, the l.h. side is non-negative, so the particle energy ${\cal E}$ has to be at least equal to ${\cal V}_{\rm eff}$ at a given place.

It is favourable that the effective potential only depends on $\hat{\nu}$ (not on $\lambda$). In the equatorial plane ($z=0$), for example, the external gravitational potential reads
\begin{equation}
  \hat{\nu}(\rho)\equiv\nu_{\rm BW}(\rho)=
 -\,\frac{2{\cal M}K\!\left(\frac{2\sqrt{\rho b}}{\rho+b}\right)}{\pi(\rho+b)}
\end{equation}
for the Bach-Weyl ring (\ref{nu,Bach}), while \citep{Semerak-03}
\begin{eqnarray}
\lefteqn{
  \hat{\nu}(\rho)\equiv\nu^{(m)}(x=0)=
  -2^{2m}(m!)^{2}\frac{2{\cal M}}{\pi\rho}
  \sum_{n=0}^{m}
  \frac{[(2n)!]^{2}(m+n)!(4n+1)}{2^{2n}(n!)^{4}(m-n)!(2m+2n+1)!}
  \;{\rm i}Q_{2n}\!\!\left({\rm i}\sqrt{\frac{b^2}{\rho^2}-1}\right)
  \;\;\;\; {\rm below~the~rim} \;\; (\rho<b) \;,} \\
\lefteqn{
  \hat{\nu}(\rho)\equiv\nu^{(m)}(y=0)=
  -2^{2m}(m!)^{2}\frac{{\cal M}}{\rho}
  \sum_{n=0}^{m}
  \frac{[(2n)!]^{2}(m+n)!(4n+1)}{2^{2n}(n!)^{4}(m-n)!(2m+2n+1)!}
  \;P_{2n}\!\!\left(\sqrt{1-\frac{b^2}{\rho^2}}\right)
  \;\;\;\; {\rm above~the~rim} \;\; (\rho>b)}
\end{eqnarray}
for the inverted $m$-th Morgan-Morgan disc (\ref{nu,iMM}) and \citep{Semerak-04}
\begin{eqnarray}
\lefteqn{
  \hat{\nu}(\rho)\equiv\nu^{(m,n)}(\rho<b)=
  -\frac{\left(1+\frac{1}{n}\right)_{m}}{n}\frac{{\cal M}}{b}
  \sum\limits_{j=0}^{\infty}
  \frac{[P_{2j}(0)]^2}{\left(\frac{2+2j}{n}\right)_{\!m+1}}
  \frac{\rho^{2j}}{b^{2j}}
  \;\;\;\;\;\; {\rm below~the~disc~rim} \;,} \\
\lefteqn{
  \hat{\nu}(\rho)\equiv\nu^{(m,n)}(\rho>b)=
  -\frac{\left(1+\frac{1}{n}\right)_{m}}{n}\frac{{\cal M}}{\rho}
  \sum\limits_{j=0}^{\infty}
  \frac{[P_{2j}(0)]^2}{\left(\frac{1-2j}{n}\right)_{\!m+1}}
  \frac{b^{2j}}{\rho^{2j}}
  \;\;\;\;\;\; {\rm above~the~disc~rim}}
\end{eqnarray}
for the ``$(n,m)$-th" power-law-density disc ($n$ is even) (\ref{nu(rho,z),z<b},\ref{nu(rho,z),z>b,even-n}).
Illustrations of the equatorial effective potential for the fields including Bach-Weyl rings and power-law discs are given in figure \ref{BH-ringdisc-Veff-equat}. It is seen that if the external source is sufficiently heavy and compact, there may exist more than one family of unstable and/or stable circular orbits --- the orbits going around the black hole alone (staying below the ring/disc) and those skirting both sources. However, the equatorial motion being integrable, its potential provides little clue as to the chaoticity of general geodesics.

The behaviour of the effective potential in the perpendicular, meridional plane can be sketched by drawing there its contour lines,
$({\cal V}_{\rm eff})^2(r,\theta;\ell)={\rm const}$.
In figure \ref{BH-ring-Veff-merid}, this is done for the angular momenta $\ell=4M$ and $\ell=0$.

The potentials simplify most notably on the axis. That of the Bach-Weyl ring reduces just to
\begin{equation}
  \hat{\nu}(z)=-\frac{{\cal M}}{\sqrt{z^2+b^2}} \;,
\end{equation}
while the inverted $m$-th Morgan-Morgan disc yields \citep{Semerak-03}
\begin{equation}
  \hat{\nu}(z)=
 -\frac{2^{2m+1}(m!)^2}{\pi}\frac{{\cal M}}{|z|}
  \sum_{n=0}^{m}C^{(m)}_{2n}{\rm i}Q_{2n}\!\!\left(\frac{{\rm i}b}{|z|}\right)
\end{equation}
and the $(n,m)$-th power-law-density disc \citep{Semerak-04}
\begin{equation}
  \hat{\nu}(z)=
 -\frac{\left(1+\frac{1}{n}\right)_{m}}{m!}
  \frac{{\cal M}}{\sqrt{z^2+b^2}}
  \sum_{k=0}^{m}\frac{(-1)^k}{kn+2}{m\choose k}
 {_2F_1}\!\left(\frac{1}{2}\,,1\,;2+\frac{kn}{2}\,;\frac{1}{1+b^2/z^2}\right),
\end{equation}
where $_2F_1$ is the Gauss hypergeometric function.
The pure axial effective potential (involving $\ell=0$ necessarily) is not very interesting, however, it always grows monotonously to zero when receding from the black hole.

Let us remark that the requirement of stationarity (or even staticity) of self-gravitating sources typically leads to the occurrence of supporting stresses within them. As opposed to the Newton's theory, these stresses contribute to gravity in general relativity, which may result in peculiar structure of space-time near the sources. This is most noticeable in the vicinity of most singular sources. Indeed, the direction-dependent effect of ring sources (even including apparent repulsive features) was noticed on geodesics by \cite{SemerakZZ-99} and further explored by \cite{D'AfonsecaLO-05}.

\subsection{Astrophysical motions in black-hole \& ring/disc fields}

Long-term ring/disc structures probably exist around black holes in X-ray binaries and in galactic nuclei. These accretion configurations are usually modelled within {\em test} approximation, i.e. their own gravitational effect is neglected. In a ``relativistic" region close to the compact centre, this is almost always apropos as far as {\em intensity} is concerned, but the additional matter may alter higher derivatives of the field which --- vice versa --- determine stability of its configuration. Actually, the most important parameters of the accretion disc like its position and extension may thus be very sensitive to the precise shape of the field which it itself generates. (This issue nicely illustrates the frequently stressed feedback between matter and its field.)

In stellar binaries, however, the accretion structures are supposed to be rather light and the whole accretion system is quite strongly (and non-stationarily) perturbed by the binary companion. Hence, the study of self-gravitating rings or discs --- and of the respective perturbed dynamics --- seems to have more relevance for the galactic nuclei. There the black holes are surrounded, in addition to inner accretion discs, by central star cluster and by massive cold tori on larger radius. The behaviour of central stars e.g. in our Galaxy is not yet fully understood and it is natural to ask about the influence of the other elements on their dynamics. In the first approximation, the torus can be described by a potential of a thin ring, the inner disc by that of an infinitely thin disc and, possibly, the cluster by an additional spherical term. In the present paper, we will thus be interested in motions of free test particles (``stars") below and above the radius of the ring/disc and monitor, in particular, whether their orbital parameters cannot change significantly over the long time spans. When the ``star" passes through the equatorial plane above the disc's inner radius, it would be also affected by the disc {\em mechanically}, but we do not try to model this interaction here (only {\em gravitational} effect is taken into account). This complements the approach of \cite{VokrouhlickyK-98} who considered Newtonian description, while {\em taking} the mechanical interaction into account.

Let us stress, however, that we try to get a first idea about the system here, not to model any actual astrophysical situation. Namely, in scanning through the phase space, the ranges of parameters are not necessarily limited to ``realistic" values.

\section{Poincar\'e sections}

\begin{figure}
\includegraphics[width=\textwidth]{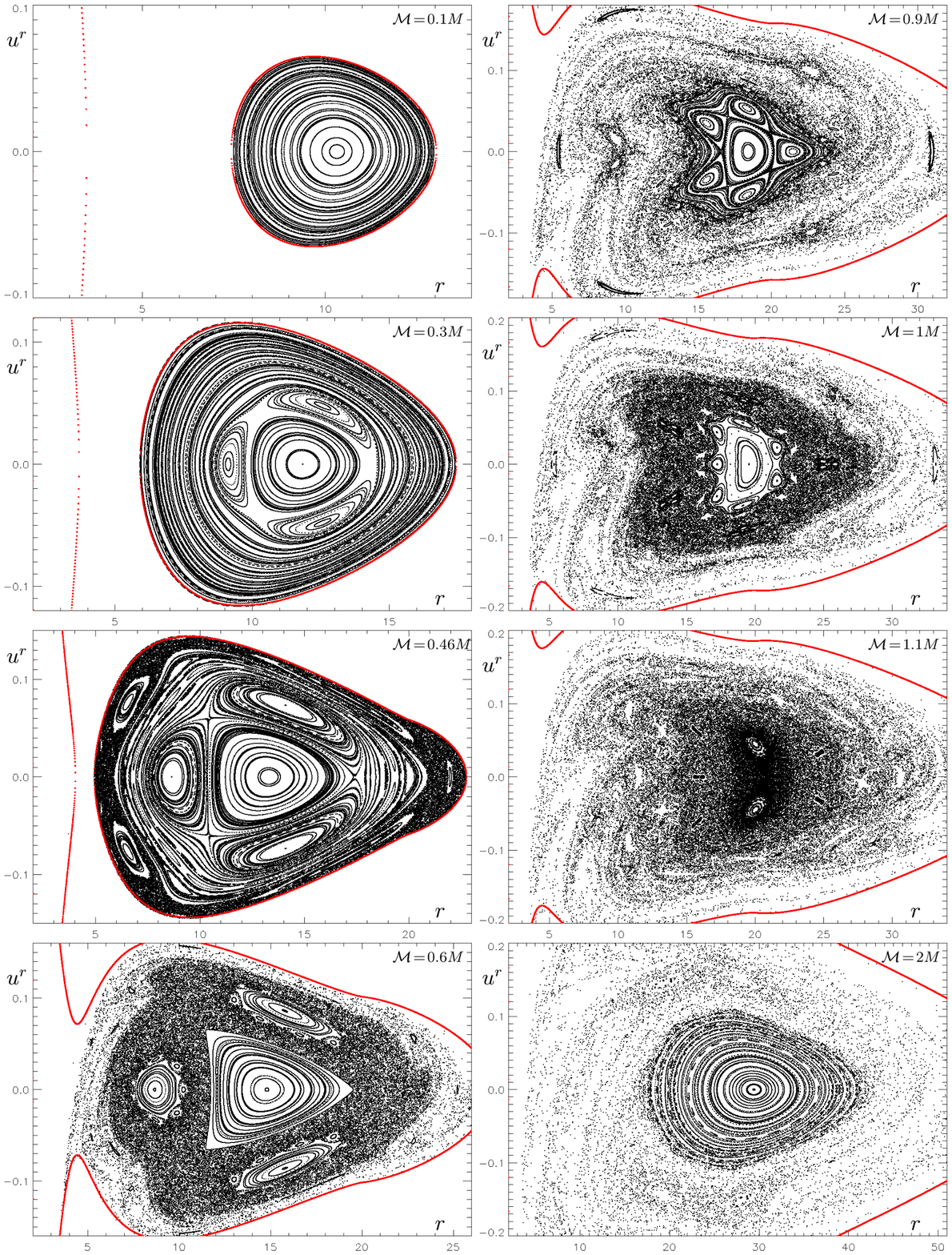}
\end{figure}

\begin{figure}
\includegraphics[width=\textwidth]{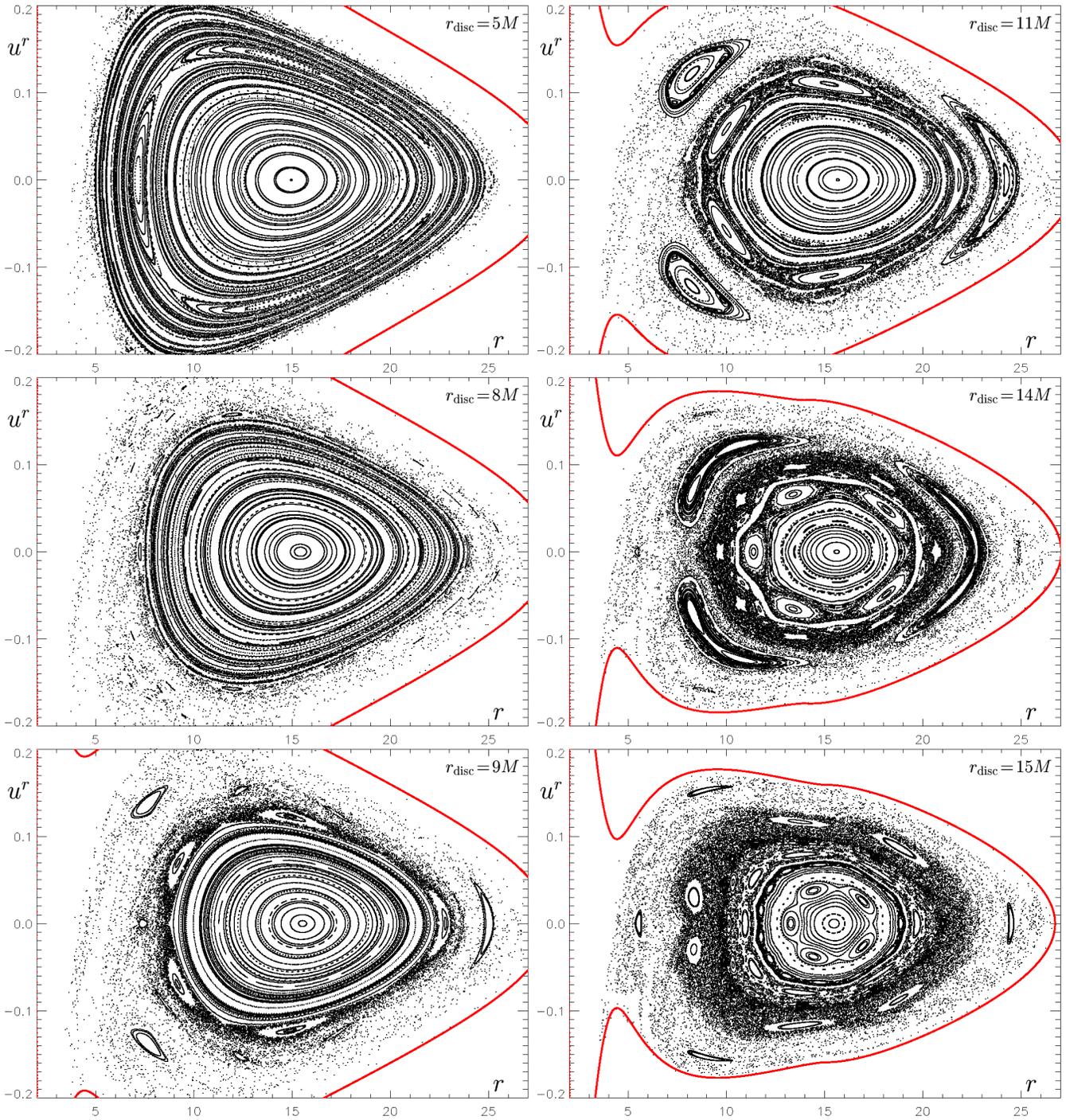}
\caption
{(The figure is on previous page!)
Geodesic dynamics in the field of a Schwarzschild black hole surrounded by the inverted 1st Morgan-Morgan disc (with radius $r_{\rm disc}=20M$): dependence on relative disc mass ${\cal M}/M$ (its value is indicated in the plots). Passages of orbits with $\ell=3.75M$, ${\cal E}=0.955$ through the equatorial plane are drawn. The boundary of the accessible region of phase space is indicated in red (this applies to all figures). \label{iMM1-m}}
\caption
{(This page.) Geodesic dynamics in the field of a Schwarzschild black hole surrounded by the inverted 1st Morgan-Morgan disc (with mass ${\cal M}=0.5M$): dependence on disc inner radius $r_{\rm disc}$ (its value is indicated in the plots). Passages of orbits with $\ell=3.75M$, ${\cal E}=0.955$ through the equatorial plane are drawn. The figure continues on the next page.}
\label{iMM1-r}
\end{figure}

\begin{figure}
\includegraphics[width=\textwidth]{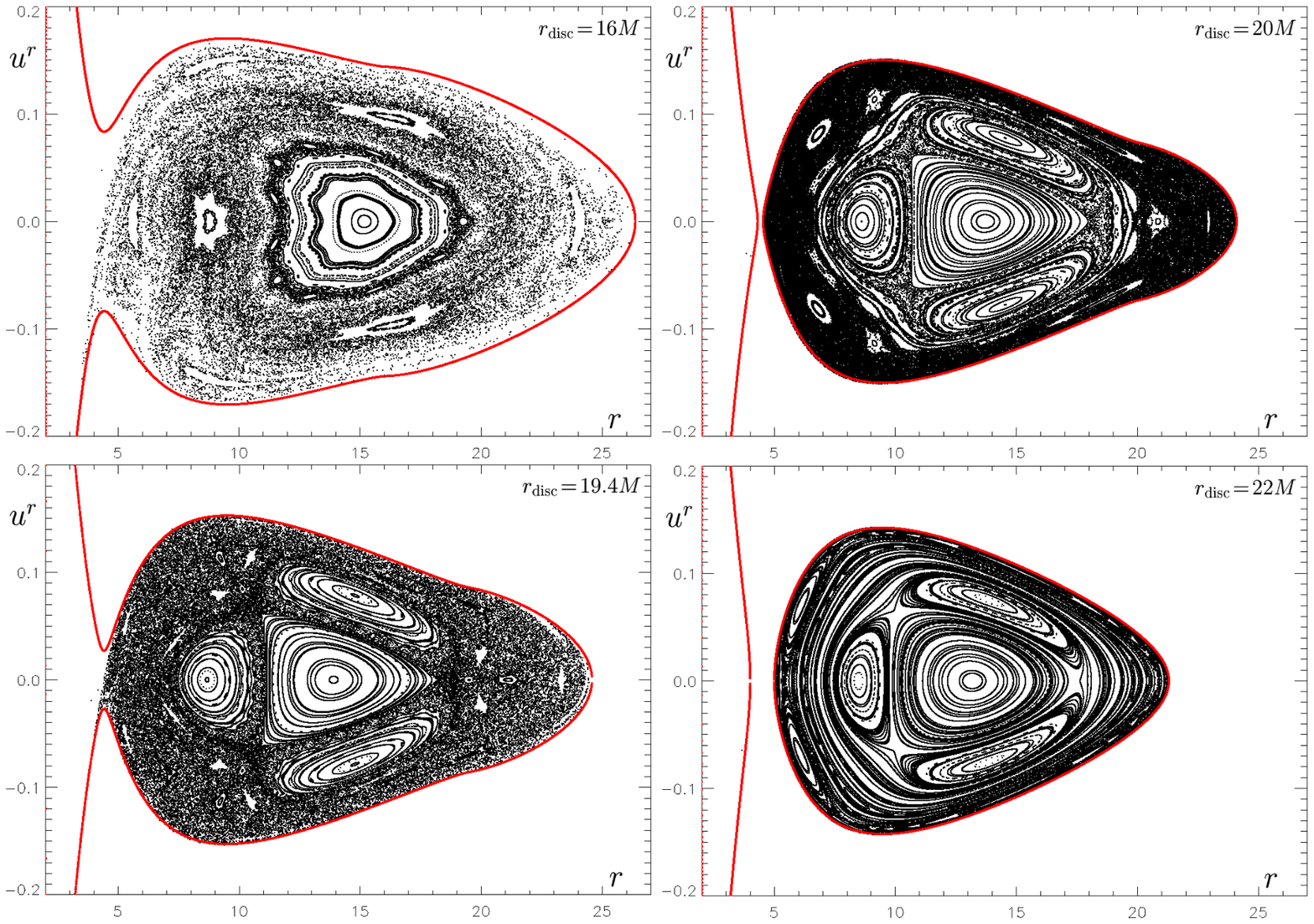}
\end{figure}

\begin{figure}
\includegraphics[width=\textwidth]{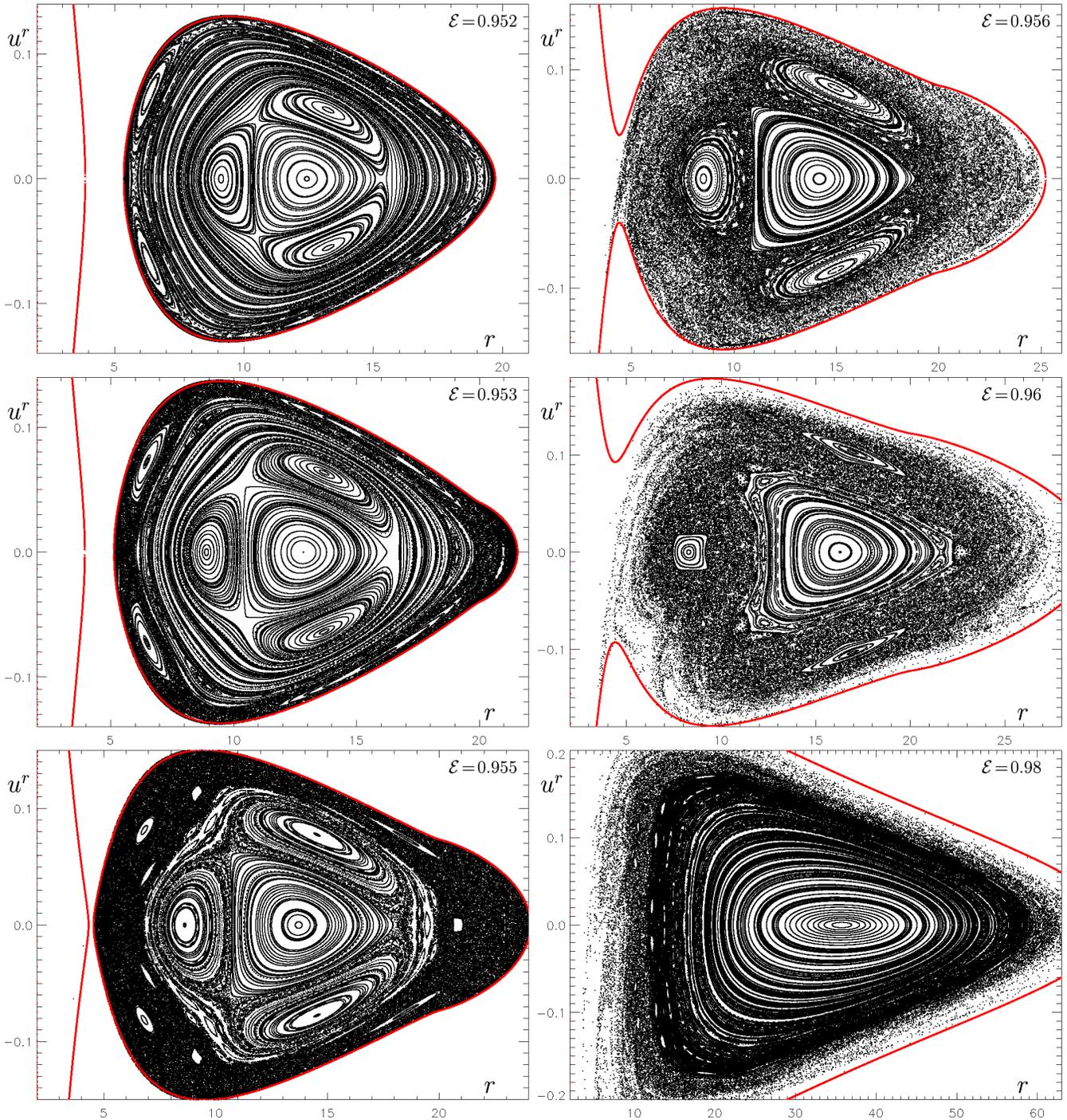}
\caption
{Geodesic dynamics in the field of a Schwarzschild black hole surrounded by the inverted 1st Morgan-Morgan disc (with ${\cal M}=0.5M$, $r_{\rm disc}=20M$): dependence on orbital energy at infinity ${\cal E}$ (its value is indicated in the plots). Passages of orbits with $\ell=3.75M$ through the equatorial plane are drawn.}
\label{iMM1-E}
\end{figure}

\begin{figure}
\includegraphics[width=\textwidth]{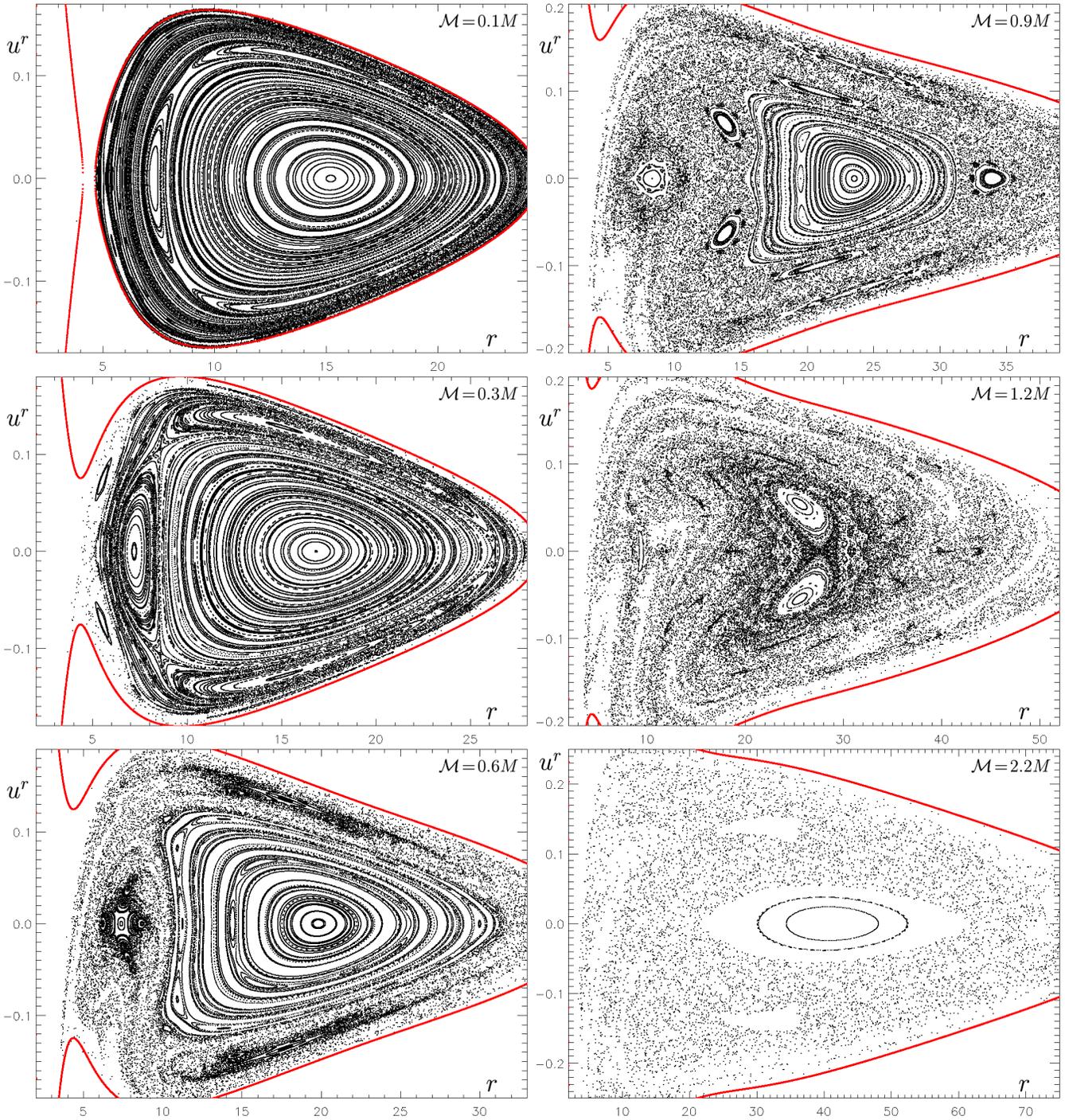}
\caption
{Geodesic dynamics in the field of a Schwarzschild black hole surrounded by the inverted 4th Morgan-Morgan disc (with $r_{\rm disc}=20M$): dependence on relative disc mass ${\cal M}/M$ (its value is indicated in the plots). Passages of orbits with $\ell=3.75M$, ${\cal E}=0.968$ through the equatorial plane are drawn.}
\label{iMM4-m}
\end{figure}

\begin{figure}
\includegraphics[width=\textwidth]{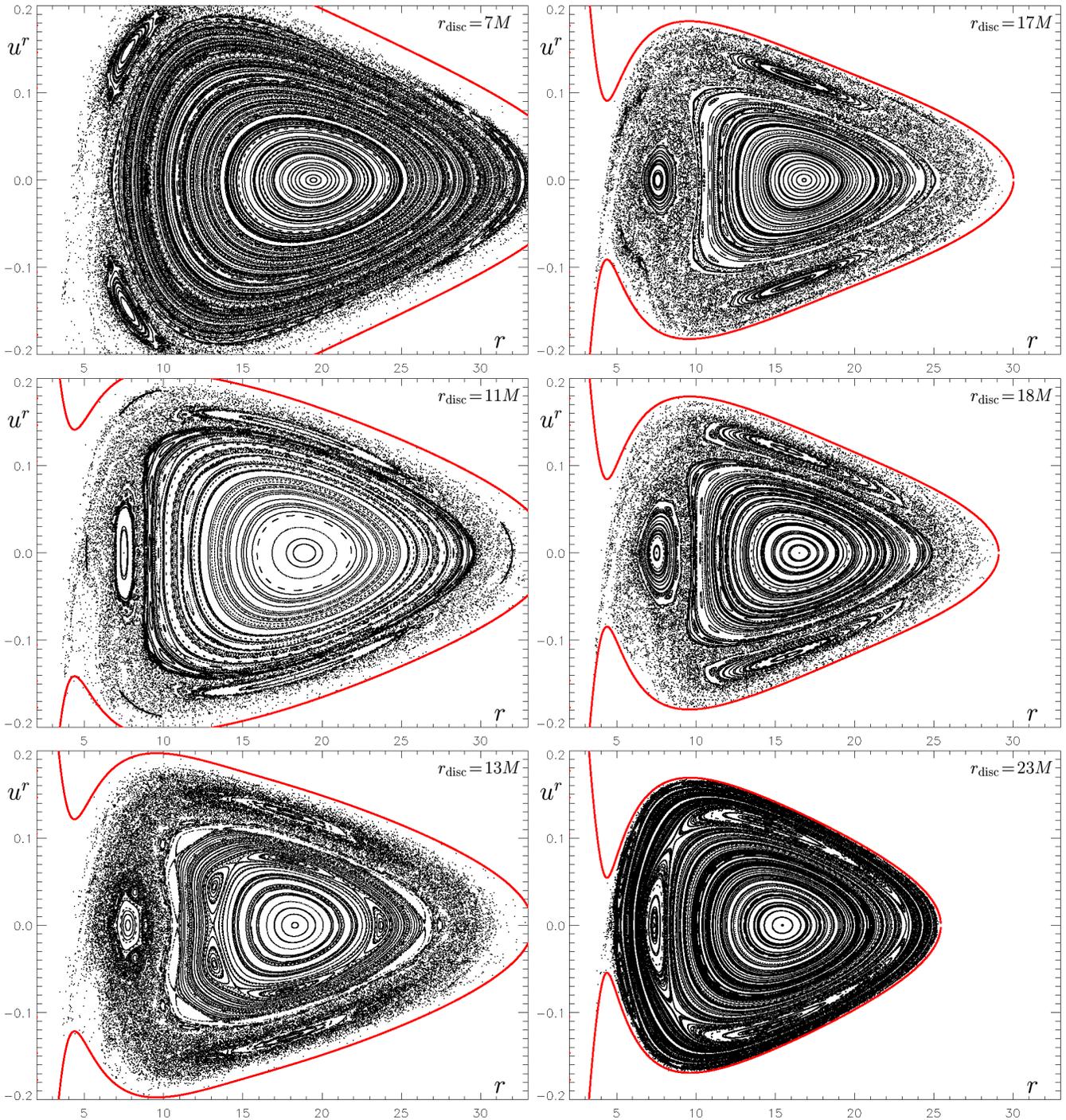}
\caption
{Geodesic dynamics in the field of a Schwarzschild black hole surrounded by the inverted 4th Morgan-Morgan disc (with ${\cal M}=0.5M$): dependence on disc inner radius $r_{\rm disc}$ (its value is indicated in the plots). Passages of orbits with $\ell=3.75M$, ${\cal E}=0.964$ through the equatorial plane are drawn.}
\label{iMM4-r}
\end{figure}

\begin{figure}
\includegraphics[width=\textwidth]{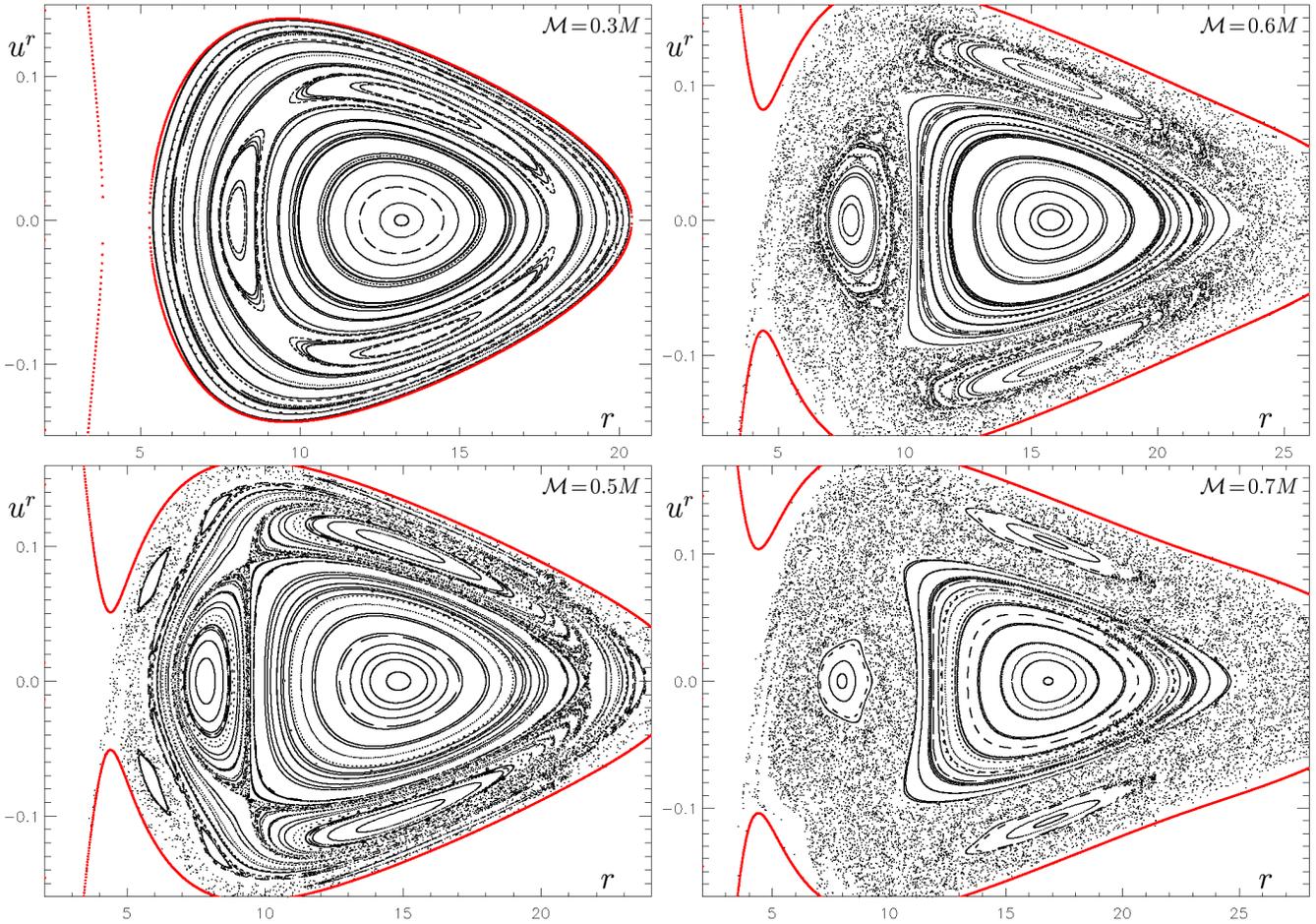}
\caption
{Geodesic dynamics in the field of a Schwarzschild black hole surrounded by the ($n$=4,$\,m$=4) power-law-density disc (with $r_{\rm disc}=20M$): dependence on relative disc mass ${\cal M}/M$ (its value is indicated in the plots). Passages of orbits with $\ell=3.75M$, ${\cal E}=0.96$ through the equatorial plane are drawn.}
\label{PL-m}
\end{figure}

\begin{figure}
\includegraphics[width=\textwidth]{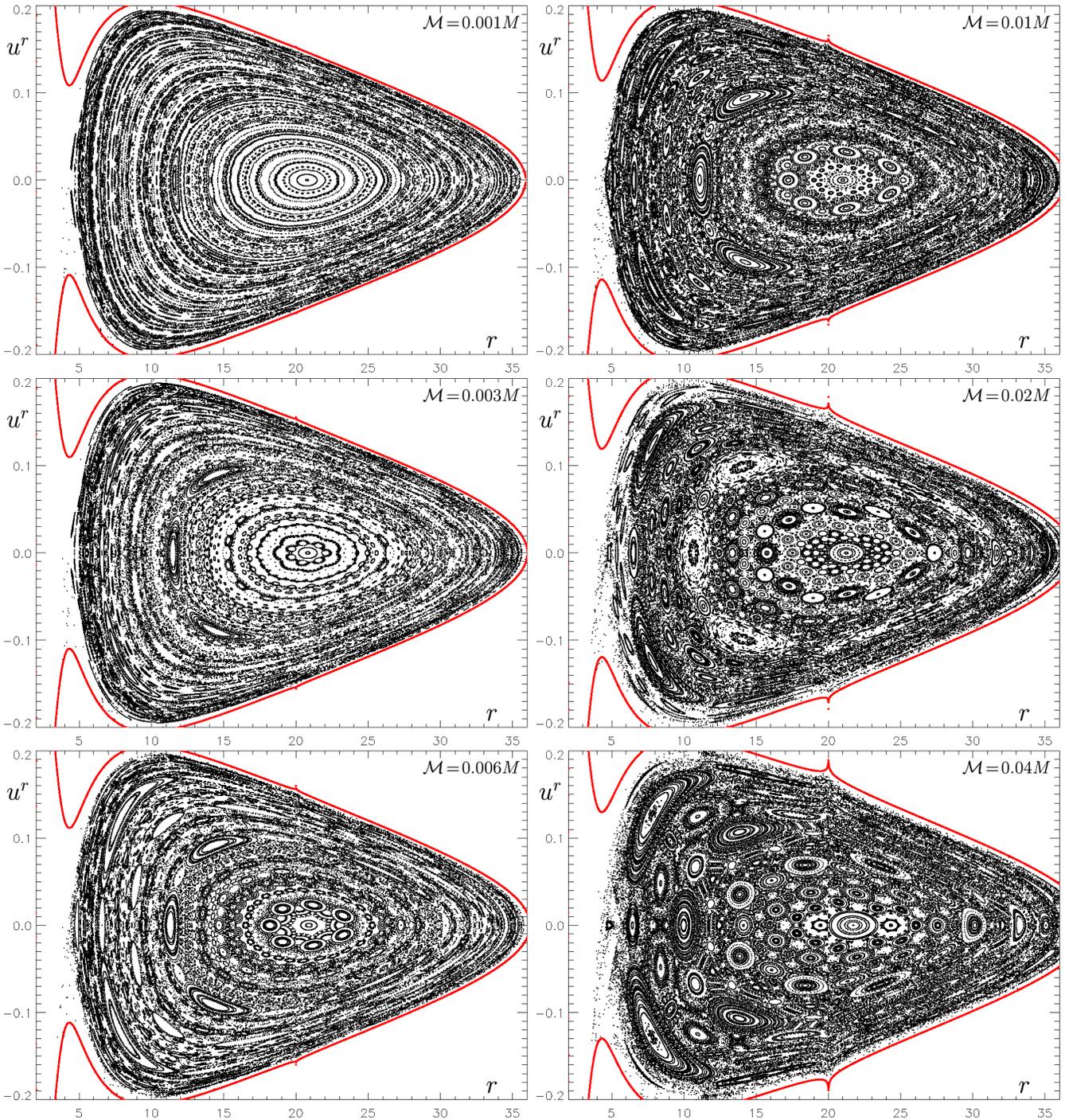}
\caption
{Geodesic dynamics in the field of a Schwarzschild black hole surrounded by the Bach-Weyl ring (with radius $r_{\rm ring}=20M$): dependence on relative ring mass ${\cal M}/M$ (its value is indicated in the plots). Passages of orbits with $\ell=3.75M$, ${\cal E}=0.977$ through the equatorial plane are drawn. The figure continues on the next page.}
\label{BW-m}
\end{figure}

\begin{figure}
\includegraphics[width=\textwidth]{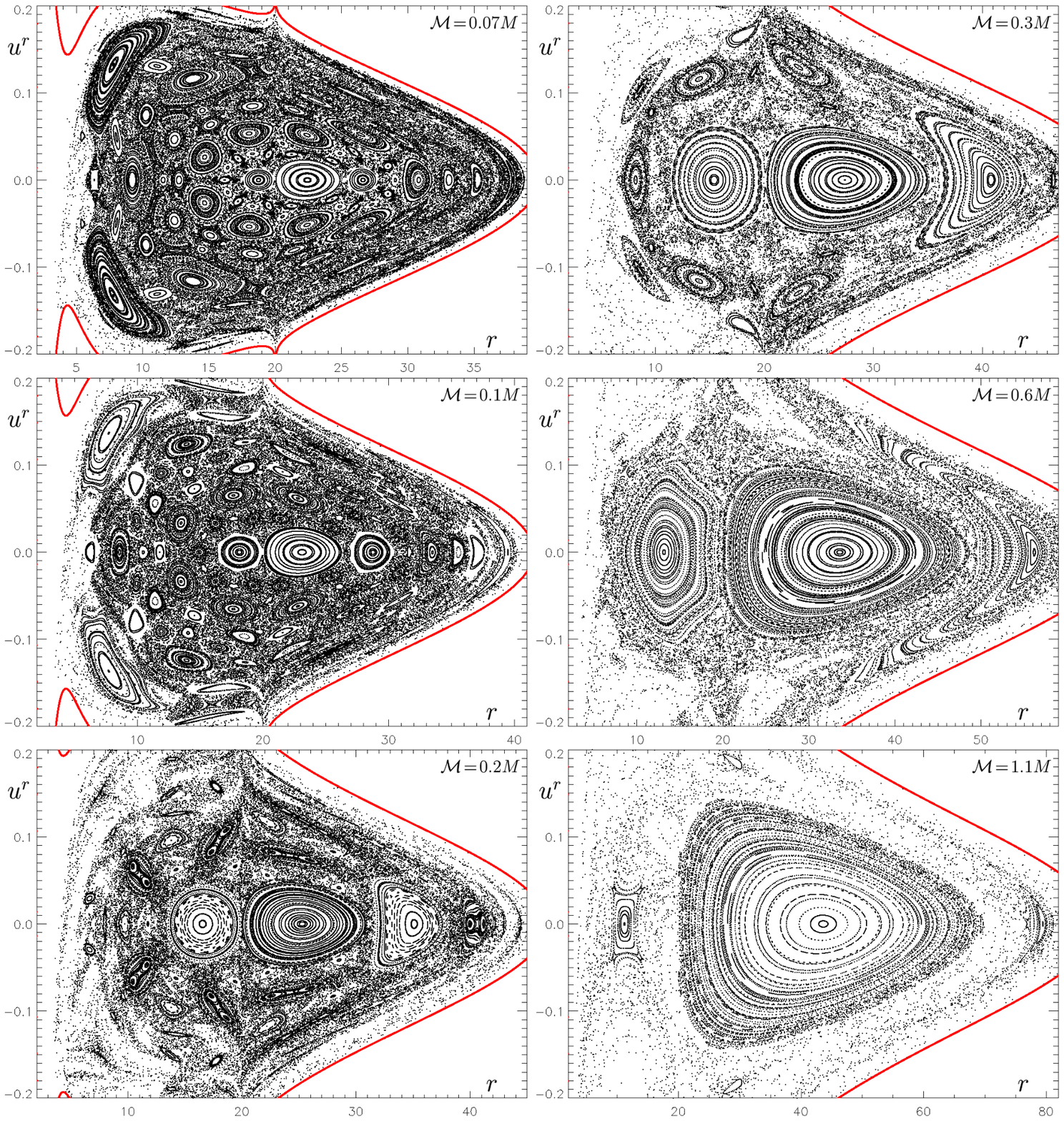}
\end{figure}

\begin{figure}
\includegraphics[width=\textwidth]{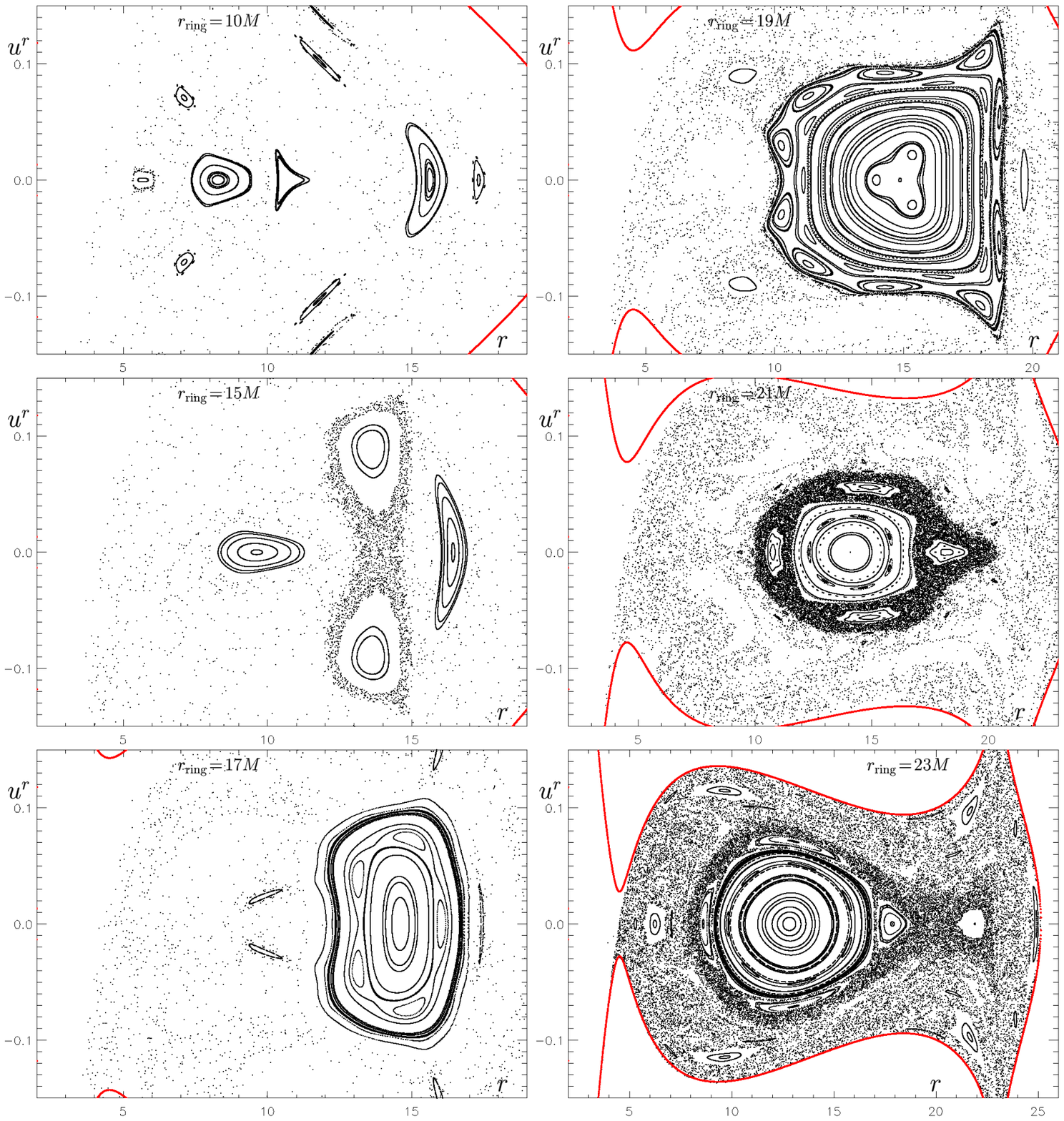}
\caption
{Geodesic dynamics in the field of a Schwarzschild black hole surrounded by the Bach-Weyl ring (with mass ${\cal M}=0.5M$): dependence on ring radius $r_{\rm ring}$ (its value is indicated in the plots). Passages of orbits with $\ell=3.75M$, ${\cal E}=0.94$ through the equatorial plane are drawn. The figure continues on the next page.}
\label{BW-r}
\end{figure}

\begin{figure}
\includegraphics[width=\textwidth]{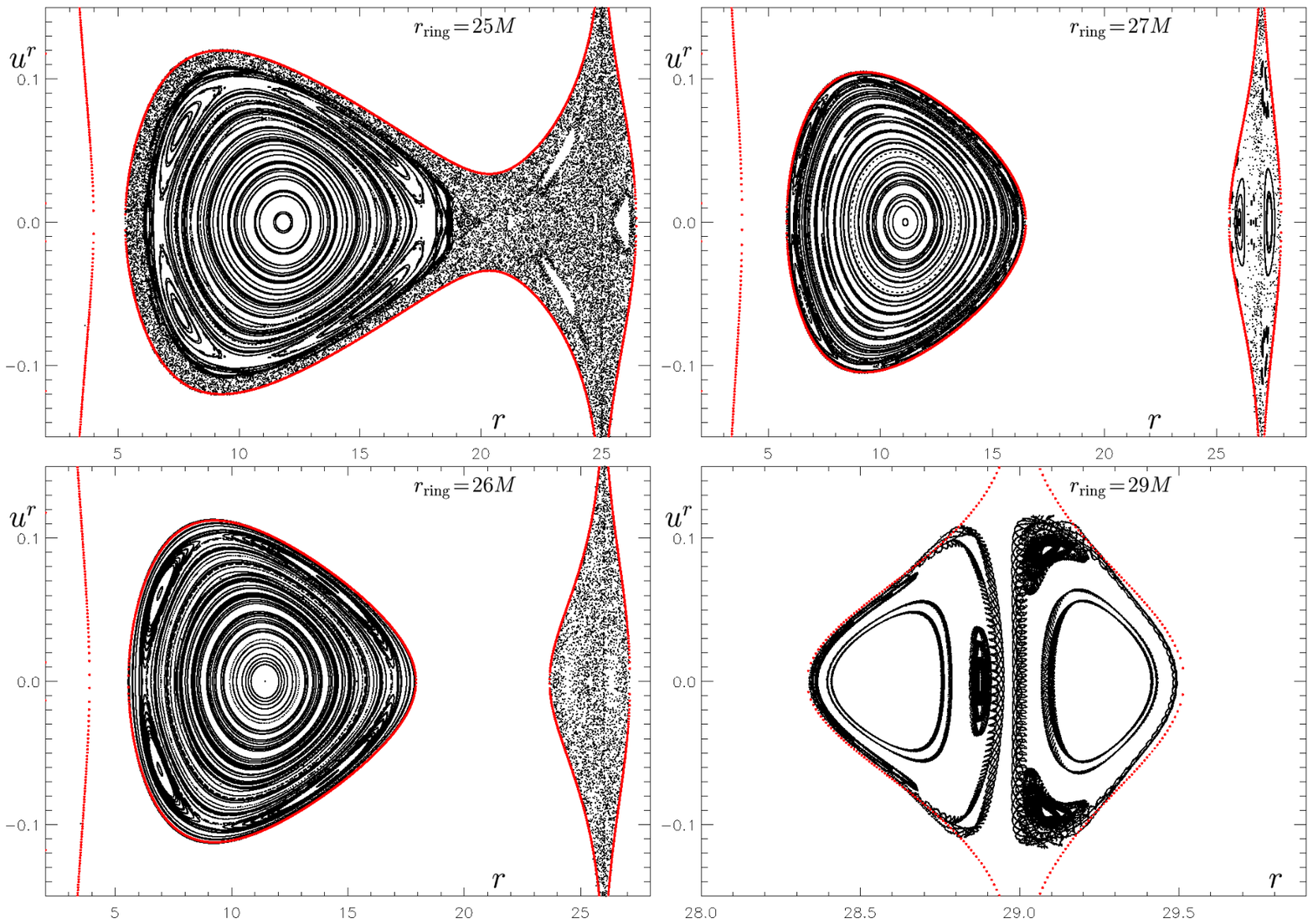}
\end{figure}

\begin{figure}
\includegraphics[width=\textwidth]{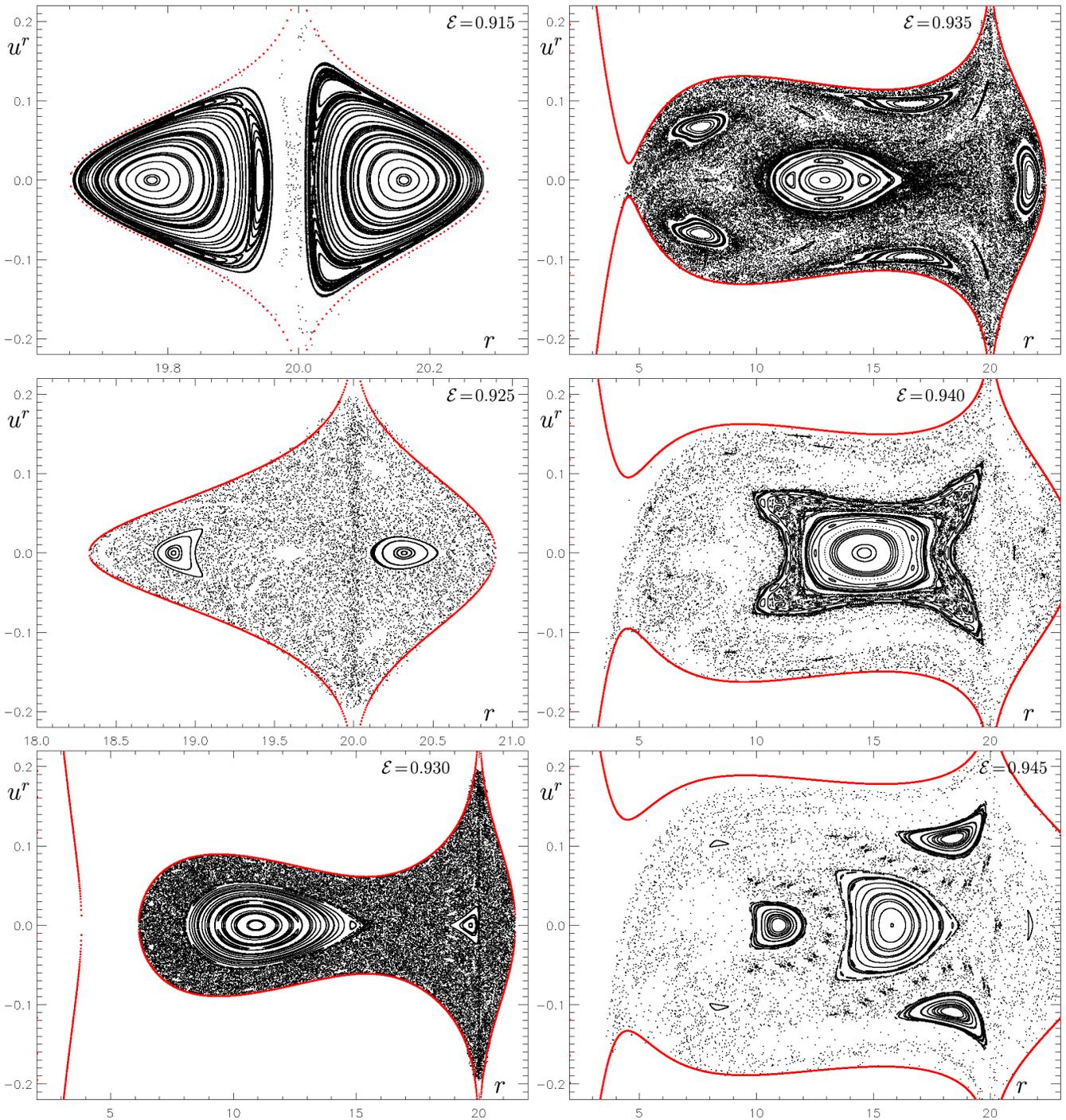}
\caption
{Geodesic dynamics in the field of a Schwarzschild black hole surrounded by the Bach-Weyl ring (with ${\cal M}=0.5M$, $r_{\rm ring}=20M$): dependence on orbital energy at infinity ${\cal E}$ (its value is indicated in the plots). Passages of orbits with $\ell=3.75M$ through the equatorial plane are drawn. The figure continues on the next page.}
\label{BW-E}
\end{figure}

\begin{figure}
\includegraphics[width=\textwidth]{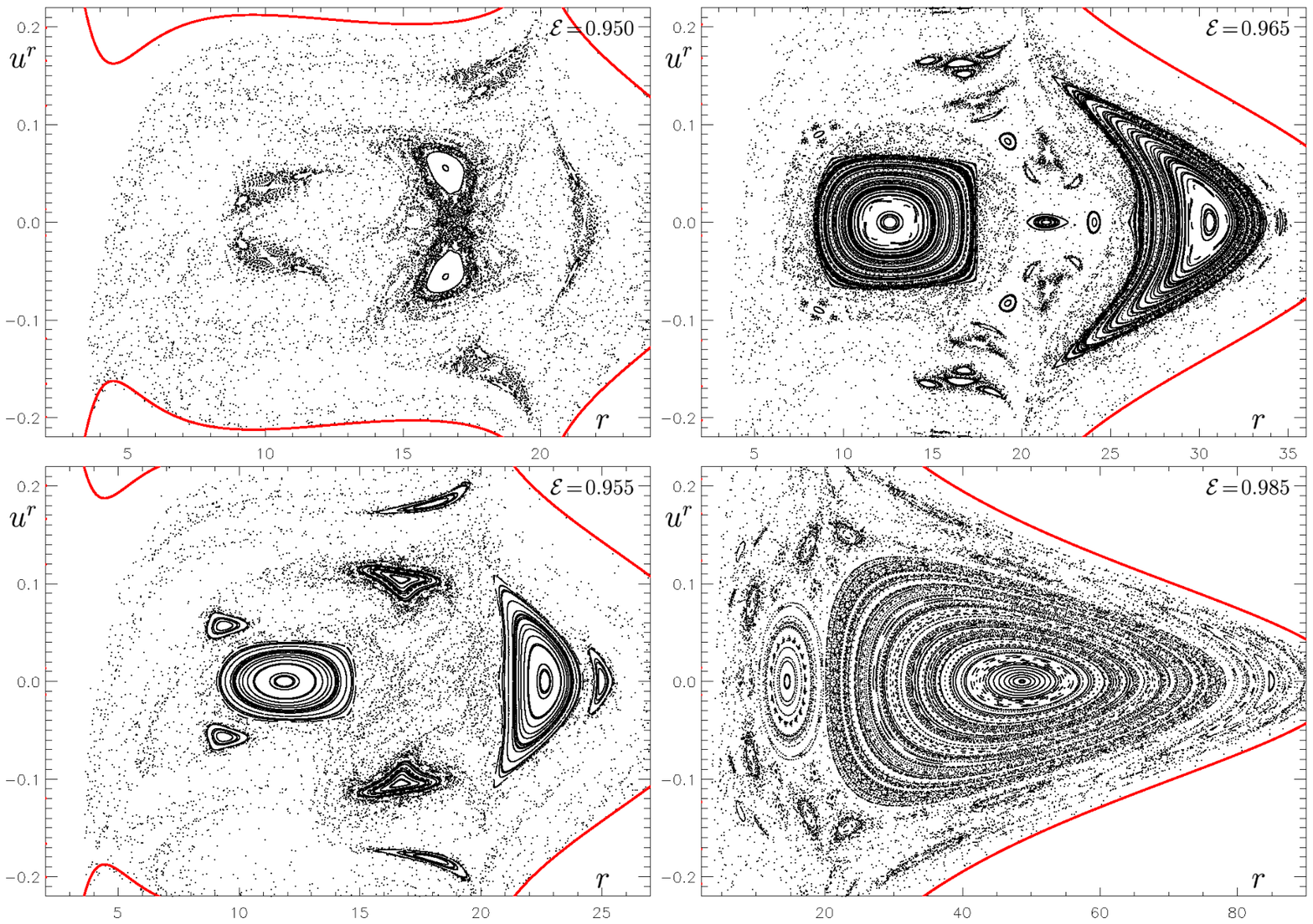}
\end{figure}

It can be expected that the originally regular, Schwarzschild geodesic dynamics will gradually grow chaotic when increasing the external-source (ring or disc) mass. We will begin the study of the system by plotting the Poincar\'e sections. Actually, the ($t$,$\phi$;$u^t$,$u^\phi$) phase subspace being completely regular due to the conservation of $u_t$ and $u_\phi$, the motion is only ``non-trivial" in the ($\rho$,$z$;$u^\rho$,$u^z$) subspace. Because of the four-velocity normalisation
\[-1=g^{\alpha\beta}u_\alpha u_\beta
    =-e^{-2\nu}{\cal E}^2+e^{2\nu}\frac{\ell^2}{\rho^2}
     +e^{2\lambda-2\nu}\left[(u^\rho)^2+(u^z)^2\right],\]
it is confined to a three-dimensional hypersurface there. This corresponds to a system with 2D Hamiltonian. In such a case, the phase-space layers with regular and stochastic motion neighbour with each other (along KAM surfaces) and Poincar\'e sections are appropriate and well tried method of how to study their arrangement. (In higher-dimensional systems the stochastic layers form a dense web that approaches arbitrarily close to any regular orbit.)

In a stationary, axially symmetric and reflectionally symmetric field, it is natural to choose the equatorial plane ($z=0$) as section on which the passages of orbits will be recorded (namely the radius and the radial component of four-velocity are usually registered). Orbits with $\ell=0$ may also be registered on the symmetry axis. Each individual Poincar\'e section will show passages of a set of particles with given energy and azimuthal angular momentum (given ${\cal E}$ and $\ell$).\footnote
{When deciding the way of choosing the initial conditions, the question always arises of what trajectories ``go together". This surely depends on the problem solved, but one usually has two major possibilities of how to define the given set of trajectories --- by fixing either their ``local", or ``global" parameters. When studying space-time motion, the ``local" choice (launching the particles with given velocity in various directions, as measured by a local physical observer) usually feels more physical, because it more closely resembles the actual astrophysical processes and does not rely on quantities referring to conditions that might nowhere occur along the trajectories (e.g. like referring to ``infinity"). However, the problem of regular/chaotic dynamics is tightly connected with the existence and values of integrals of motion, so we will adhere to these ``global" quantities in assigning the initial conditions. Fixing such ``global" parameters also more probably confines the orbits to similar regions in phase space, which makes the comparison of their long-term evolution easier.}

The existence and nature of the constants of motion imply that certain subsets of geodesics will remain regular, whether the additional source is present or not. Specifically, ${\cal E}$ is purely connected with $u^t$ and $\ell$ with $u^\phi$, plus there is the four-velocity normalisation, so the regularity applies to the geodesics lying fully within the 3-dimensional world-sheets spanned by $t$, $\phi$ and one of the meridional coordinates. For our systems this factually means the geodesics bound to the equatorial plane ($z=0$, $u^z=0$): due to the reflectional symmetry, $u^z=0$ is their fourth integral of motion.

We will study, on Poincar\'e sections, the change of time-like geodesic structure in dependence (i) on relative mass and position of the external ring/disc and (ii) on constants of the geodesic motion (namely energy; the dependence on angular momentum is less intriguing). The geodesics are computed using the (adjusted) programme written by M. \v{Z}\'a\v{c}ek for our earlier paper \citep{SemerakZZ-99} where the motion of free test particles was followed in similar fields we deal with here. In all figures, a set of several hundreds of particles is launched from a certain region of relevant radii in the equatorial plane, with given ${\cal E}$ and $\ell$ (which determine $u^t$ and $u^\phi$, respectively), with several different values of $u^r$ within the local light cone and $u^\theta$ components fixed accordingly from the four-velocity normalisation. In most plots, the particles are followed for about $150000M-250000M$ of proper time; this usually corresponds to some $200-1000$ orbital periods, i.e. to $400-2000$ recorded points (since our section $z=0$ is the plane of symmetry, we can register passages in both directions).

Let us note that it is no point to specify the initial conditions and evolution of individual orbits precisely, because the figures cannot be compared ``invariantly" anyway: they correspond to different space-time backgrounds (this applies to sequences showing dependence on ${\cal M}$ and on $r_{\rm in}$) or at least to different regions of permitted $u^r(r)$ (as in sequences showing dependence on ${\cal E}$). This also implies that the plots naturally do not represent the same number of particles and neither the same average ``density" of crossings. In particular, those with the permitted ($r$,$u^r$)-region opened towards the horizon tend to be more ``diluted" due to the escape of particles into a black hole.

The geodesic dynamics of perturbed systems is usually rather complicated and one has to look through many Poincar\'e sections in order to catch its evolution with the parameters. Even single sections are often quite rich and require enough space in order to be represented properly. We hope the following groupings of plots are a reasonable compromise between providing accurate information and keeping the paper tolerably long.

Figures \ref{iMM1-m}, \ref{iMM1-r} and \ref{iMM1-E} concern the space-time of a Schwarzschild black hole surrounded by the inverted first Morgan-Morgan disc; they illustrate the dependence of geodesic dynamics on relative mass of the disc ${\rm M}/M$, on its inner radius $r_{\rm disc}$ and on energy of the geodesics ${\cal E}$, respectively.
For comparison with different discs, figures \ref{iMM4-m} and \ref{iMM4-r} show the dependence on ${\rm M}/M$ and $r_{\rm disc}$ for the Schwarzschild black hole surrounded by the inverted fourth Morgan-Morgan disc, and figure \ref{PL-m} shows the dependence on ${\rm M}/M$ for the hole with the ($n$=4,$\,m$=4) power-law-density disc.
Finally, figures \ref{BW-m}, \ref{BW-r} and \ref{BW-E} provide the same kind of information for fields involving the most concentrated, singular external source --- the Bach-Weyl ring.
In all the figures, the Schwarzschild radius $r$ is given along the horizontal axis in units of the black-hole mass $M$. The boundary of the accessible region of phase space is indicated in red.

The presented sequences of plots can be best annotated by quotation from the introduction of the book \cite{LichtenbergL-92}. When describing the phase-space portrait of nearly-integrable systems and KAM theory, the authors write: ``The invariant surfaces break their topology near resonances to form island chains. Within these islands the topology is again broken to form yet other chains of still fines islands. On ever finer scale, one sees islands within islands. But this structure is only part of the picture, for densely interwoven within these invariant structures are thin phase space layers in which the motion is stochastic." (It should be remarked, however, that the KAM theory usually assumes continuous force field, which is not satisfied along orbits crossing the disc.)

We do not present any systematic comparison of different additional sources nor a thorough account of dependences on parameters here, the selected figures should just illustrate overall tendencies. Although they clearly contain a number of interesting details, it would be too lengthy to take notice of them individually. Therefore, let us only mention rather generic observations:

\begin{enumerate}
\item
Dependence on angular momentum.
Larger angular momentum means larger part of (some given) energy allotted to azimuthal motion. This component of motion is ``held" by the conserved value of $\ell$, however, so larger angular momentum favours regularity. Larger $\ell$ also means higher centrifugal barrier around the black hole; for a certain value of $\ell$ the barrier even prevents the particles from falling into a hole. This is advantageous for the present study since the particles then do not ``drop out" of the phase space and can be followed for a sufficient number of orbital periods. (The dependence on $\ell$ is otherwise not so interesting and we do not focus on it. When exploring the dependence on other parameters, we mostly choose $\ell=3.75M$.)
\item
Dependence on energy.
The picture begins with one regular island around an orbit with periodicity $k=1$. With increasing energy, the primary island enlarges and shifts to higher radii, smaller islands of higher-periodic orbits arise within it and stochastic layers gradually develop from separatrices around them (mainly from the latter's vertices, i.e. hyperbolic points of the KAM theory, and near the boundary of the accessible region first). Stochastic layers tend to coalesce and form a ``chaotic sea" while smaller islands of higher and higher resonances emerge (and often disappear again). At certain phase the picture is rather complicated, containing both the residues of regular islands of various sizes, narrow stochastic layers of ``almost regular" trajectories (which remain very close to some regular orbit for a long time, while only slightly changing their action), and also large ``chaotic sea" of orbits with strongly variable action (usually spread from the boundaries of the phase-space region accessible to the particles with given constants of motion). For large energies the higher-resonance islands continue to shrink towards the margin of the accessible zone, but the primary island begins to grow back and finally occupies almost all phase space again. This is also connected with the evolution of the accessible region (its boundaries are indicated in red in the figures). For small energies it typically consists of two parts, one just above the horizon and the second below (and/or around) the external source. With increasing energy, the two lobes get closer and finally join, permitting the particles to fall below the horizon. This effectively drains the chaotic particles (the most eccentric ones) from the system.
\item
Dependence on mass of the additional source (strength of the perturbation).
The behaviour is similar to the dependence on energy: with increasing relative mass, higher and higher resonances gradually appear, release into a developing chaotic sea and shrink towards the boundary of the accessible zone. The degree of chaoticity is the highest when the masses of both sources are comparable. (Even the $k=1$ orbit may become unstable and its island may break up, giving way to chaos in the central region.) When the total gravity of the external source prevails, however, the system rather inclines back to regularity --- the central ``circular" orbit again turns stable and its island gradually spreads out.
\item
Dependence on position of the additional source.
The position of the ring/disc may affect the total potential even more than the external-source mass and this also reflects on geodesics. When the source is very close to the hole, it does not ``develop" its own potential minimum with ``its own" periodic orbits (unless it is extremely massive), so the dynamics remains almost regular (the particles only feel the centre heavier). Shifting the source farther from the hole, the potential valleys with periodic orbits originate. Their islands and separatrices evolve in a complicated manner, making way to stochastic layers and arising again elsewhere. However, when the ring/disc is above some radius, the chaotic orbits begin to gather into narrow belts again, invariant curves finally emerge and form a primary and the higher islands, gradually spreading towards the margins of the accessible region.
\end{enumerate}

\noindent
Several more general points:

\begin{enumerate}
\item
More compact source perturbs the dynamics more. The Poincar\'e-section series thus reveal slower evolution with parameters for sources that are more ``diffuse". Of the discs considered here, the inverted 1st Morgan-Morgan disc is more compact than the 4th, the power-law disc ($n\!=\!4$,$\,m\!=\!4$) being quite similar to the latter. The Bach-Weyl ring is of course the most compact perturbation and deserves a separate comment. The ring being singular, it {\em always} has a certain potential valley of its own where particles may evolve without passing over to the black-hole sphere of influence. In this region surrounding the ring the phase-space structures behave similarly as we described above, in particular, the geodesic motion may become chaotic there to a large measure. At the same time, a completely regular closed region may exist around a potential minimum between the hole and the ring --- see e.g. figure \ref{BW-r}. (In our figure series the accessible lobes are mostly joined, however.) The figures show that the hole-ring phase portrait is extremely rich and very strongly dependent on the parameters. Although the overall tendencies are similar to those observed with discs, there are much more and distinct higher resonances, already occurring at lower values of the parameters. For instance, the ring of mass ${\cal M}=0.001M$ already induces a considerable perturbation of the phase pattern in almost the whole volume of the accessible zone (cf. figure \ref{BW-m}). However, no chaotic sea is formed until values around ${\cal M}=0.1M$, the numerous regular islands and chaotic layers remain densely interwoven.
\item
The dynamics does not respond to the change of perturbation parameters in a ``monotonous" (and by no way ``linear") way. Within some intervals of the parameter space just minor quantitative changes occur, whereas elsewhere the picture ``quickly" alters considerably (regularity islands suddenly disappear, sometimes arising in a totally different arrangement in a moment, indicating change in orbital periodicity. (A similar remark may even apply to one single trajectory in a one single background --- see section \ref{series-spectra} below).
\item
When increasing a given parameter, one often notices {\em several} periods of prevailing regularity and several rather stochastic ones; however, the dynamics typically inclines to regularity for {\em both very small and very large} values of the parameters. (This behaviour has also been observed in other systems --- cf. \cite{StranskyKC-06}, for example.)
\item
For our gravitational systems, the geodesic dynamics rather tends to ``break up" from the boundaries of the accessible phase-space region, while a certain regular region mostly (though not always) survives in the interior.
\item
The Poincar\'e sections are generally not symmetrical with respect to $u^r=0$. This specifically applies to those cases where the accessible region of phase space is open towards the black hole. Namely the asymmetry is due to the particles that fall into the hole (notice, in those cases, the excess of dots with $u^r<0$ at low radii); these are not necessarily balanced by counterparts that would be launched outwards from near the horizon.
\item
We also drew several axial sections $\theta=0$ for orbits with zero angular momentum, but these are not presented here.
\end{enumerate}

\subsection{A comment on numerics}

We solve the geodesic-equation system (\ref{ddt})--(\ref{ddz}) by the Runge-Kutta (actually the Hut$\!$'a) 6th-order method with variable proper-time step, on the cluster {\it Tiger} operated at the Astronomical Institute of our university.
The numerical and computational demands strongly depend on specific space-time background. The case with Bach-Weyl ring and with 1st inverted Morgan-Morgan disc can be integrated quite fast and with high precision --- constants of the motion and four-velocity normalisation are conserved there with relative error $10^{-14}$ or less. For the 4th inverted Morgan-Morgan disc the error is about an order larger. For the power-law-density disc the computation takes about 30 times longer than with the previous sources and the error is larger, about $10^{-10}$ (which is not always sufficient as seen in figures \ref{PL-m}: contours within the regular regions often get blurred there at places where the phase space is more ``dense"). This is because its potential is given by an infinite sum of Legendre polynomials, namely (\ref{nu(rho,z),z<b}) or (\ref{nu(rho,z),z>b,even-n}): the latter converge rather slowly and have to be cut somewhere. In the code we add up all terms which are greater then $10^{-13}$ in both the sums. Here we mean the sums themselves, without the coefficients in front. (For the particular disc we consider, i.e. $n=4$, $m=4$, the front coefficient $\frac{\left(1+\frac{1}{n}\right)_{m}}{n}\frac{{\cal M}}{b}$ is approximately $9.7\/{\cal M}/b$ which is typically of the order $0.01\div 1$.) For comparison: at place of the worst behaviour, $\sqrt{\rho^2+z^2}\simeq b$, the magnitude of the leading ($j=0$) term of the sums is about $0.034$. Trying to estimate the tendency of farther terms, one finds that on the axis ($\rho=0$) at $|z|=b$, the $j$-th term of the sum (\ref{nu(rho,z),z<b}) reads
$\left[\left(\frac{2+2j}{n}\right)_{m+1}\right]^{-1}\!
 \frac{(-1)^j(2j)!}{2^{2j}(j!)^2}\;$;
for $n=4$, $m=4$ this falls off as $\sim j^{-11/2}$ at large values of $j$.
(We note in passing that $388$ terms have to be summed there according to our $10^{-13}$ criterion; close to the disc, even more of them may be necessary. On the other hand, in regions farther from the $\sqrt{\rho^2+z^2}\simeq b$ interface, several tens of terms usually suffice.)

Finally, let us add that the numerical integration is not accurate close to the horizon (regardless the type of the external source): notice that some of the particles plunging into the hole got {\em beyond} the boundary of the accessible zone.

\section{Time series and their Fourier spectra}
\label{series-spectra}

\begin{figure}
\includegraphics[width=0.862\textwidth]{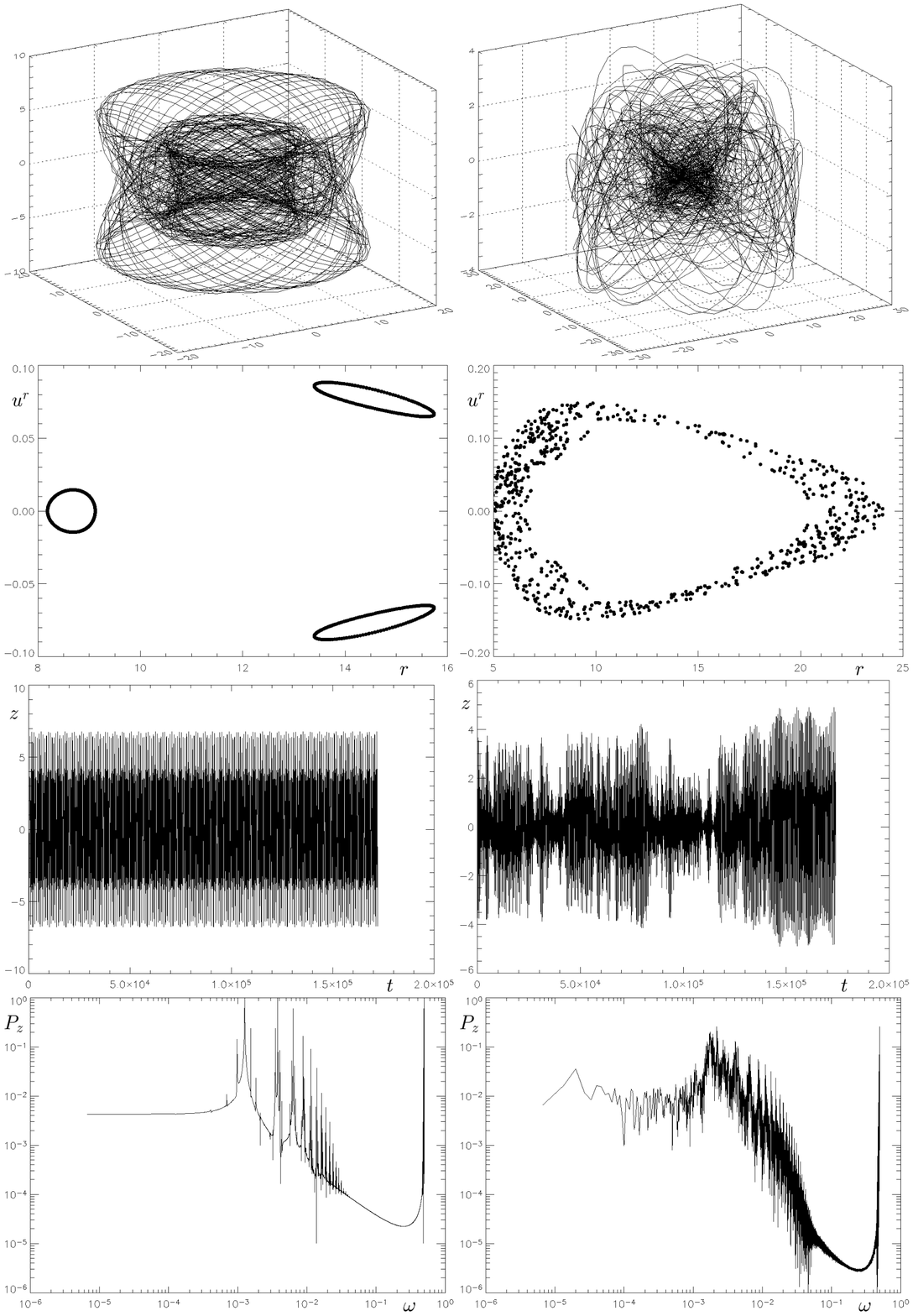}
\caption
{Comparison of a selected regular ({\it left}) and chaotic ({\it right}) orbit, both with constants ${\cal E}=0.955$, $\ell=3.75M$ and obtained in the field of a Schwarzschild black hole surrounded by the inverted 1st Morgan-Morgan disc with ${\cal M}=0.5M$, $r_{\rm disc}=20M$. From top to bottom, the rows contain spatial tracks of the orbits, their Poincar\'e sections $z=0$ ($r$,$u^r$), time series of the $z$ position and corresponding Fourier spectra. Coordinates are in the units of $M$, frequencies in the units of $1/M$.}
\label{spectra-1}
\end{figure}

\begin{figure}
\includegraphics[width=\textwidth]{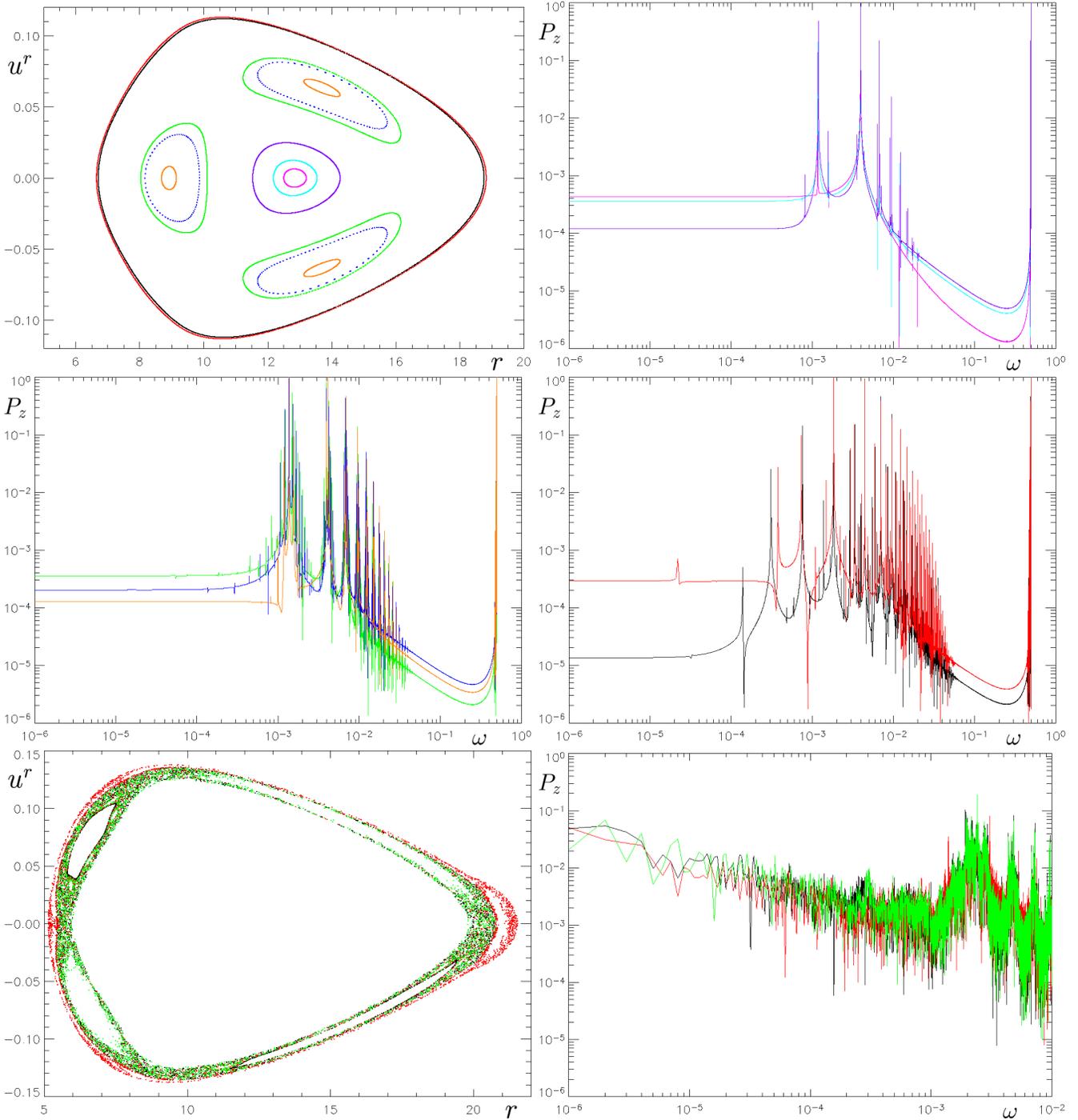}
\caption
{Comparison of selected orbits with constants ${\cal E}=0.953$, $\ell=3.75M$, obtained in the field of a Schwarzschild black hole surrounded by the inverted 1st Morgan-Morgan disc with ${\cal M}=0.5M$, $r_{\rm disc}=20M$. At top left, 3 types of regular orbits are chosen in respective Poincar\'e section. The spectra of their $z$-position time series are given at top right (three primary-island orbits), middle left (three orbits from islands of periodicity $k=3$) and middle right (two orbits skirting the whole island and neighbouring with the chaotic layer). In the bottom row, three ``weakly chaotic" orbits are chosen in the same Poincar\'e section and their spectra shown on the right. The individual orbits are given in different colours. Radius is in the units of $M$, frequency in the units of $1/M$.}
\label{spectra-2}
\end{figure}

\begin{figure}
\includegraphics[width=\textwidth]{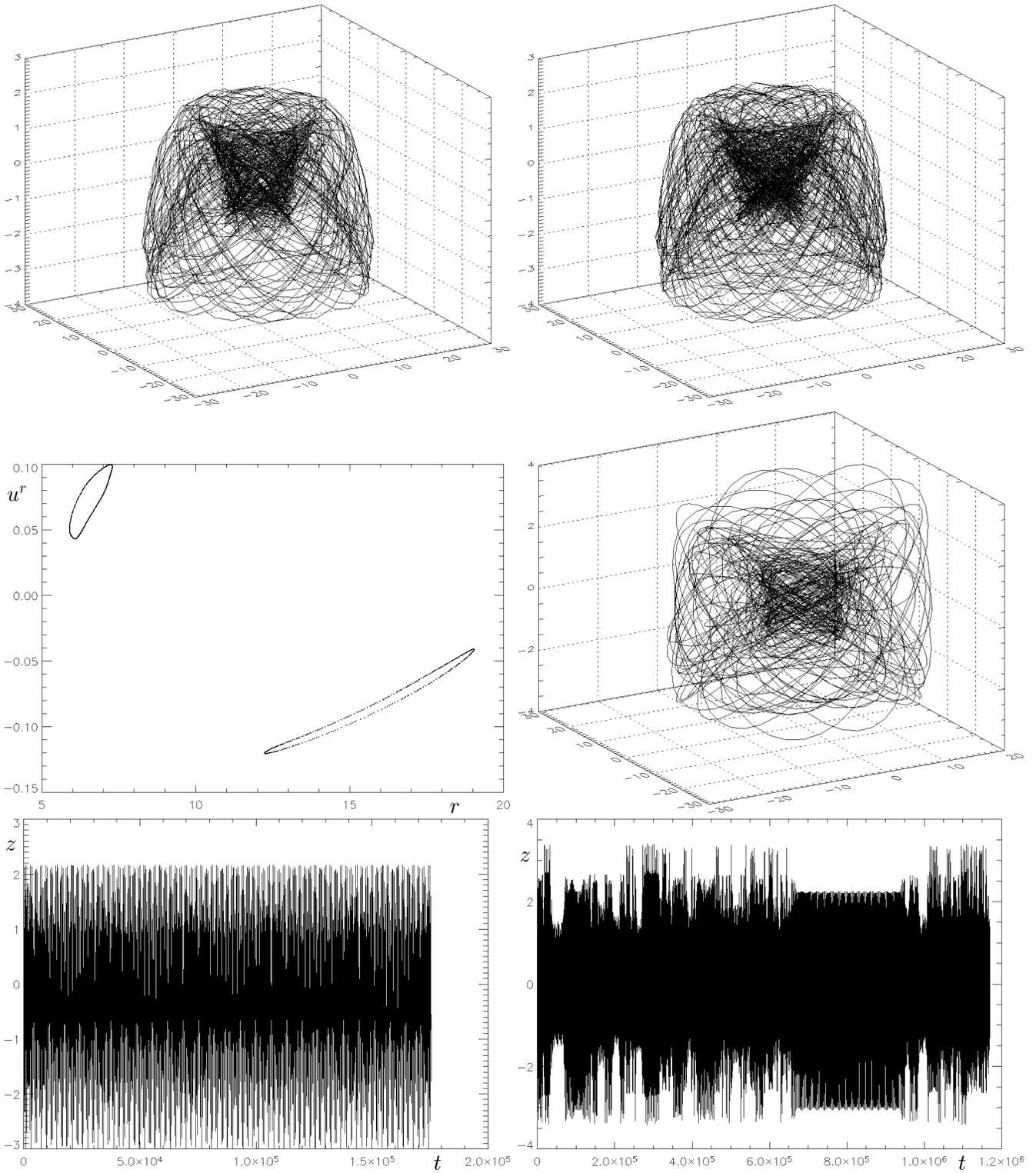}
\caption
{Further comparison of the orbit sticking to the $k=2$ islands in Fig. \ref{spectra-2} (the black one in the bottom row) and a selected regular orbit belonging to these islands. On the left, the spatial track (top), the Poincar\'e section (middle) and the time series of $z$ (bottom) are drawn for the regular orbit. On the right, the spatial tracks for the weakly chaotic orbit are drawn, at ``almost regular" times $7\cdot 10^5M<\tau<9\cdot 10^5M$ (top) and at more chaotic times $\tau<100000M$ (middle); the time series of $z$ is given at the bottom.}
\label{spectra-3}
\end{figure}

\begin{figure}
\includegraphics[width=\textwidth]{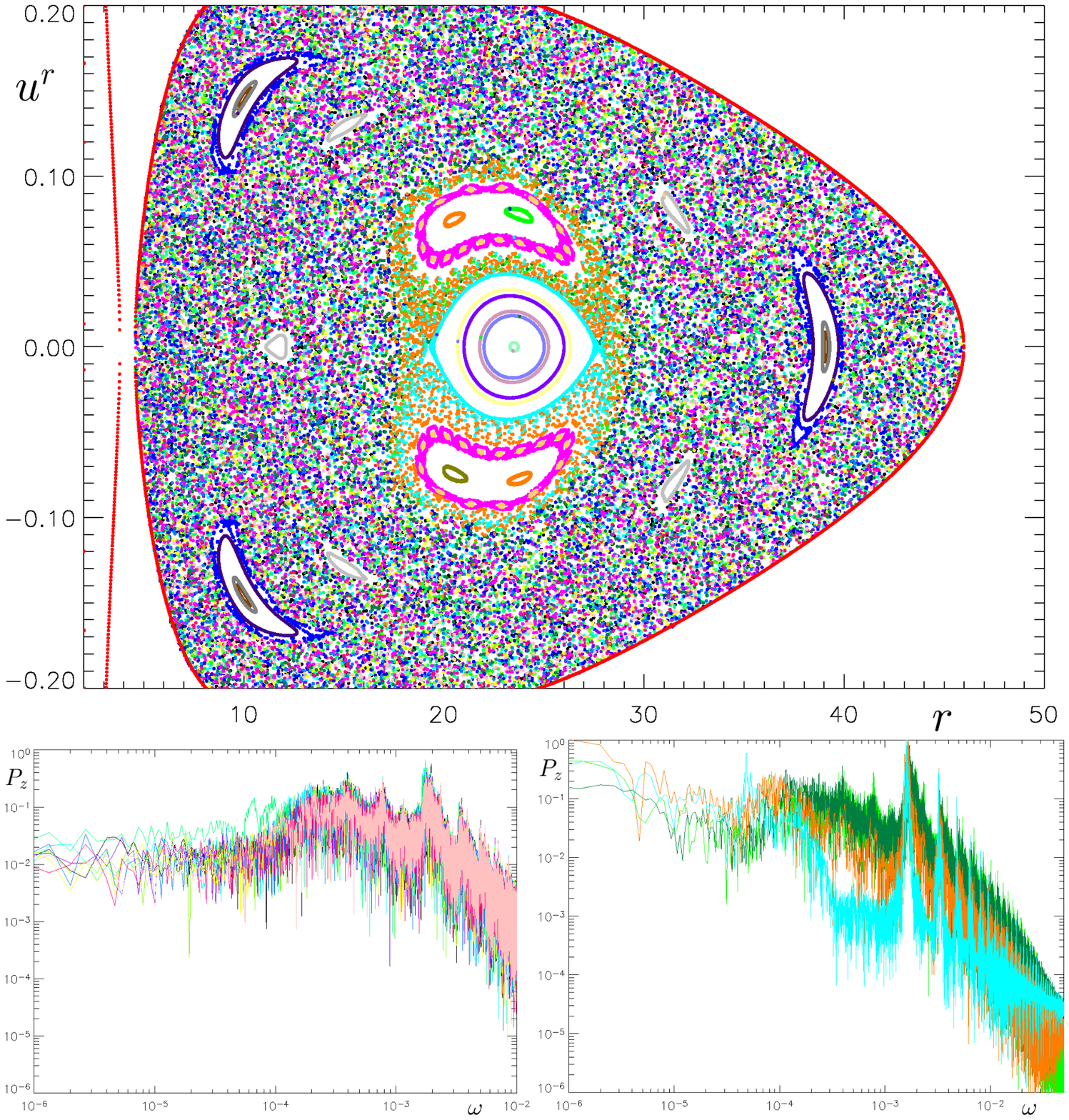}
\caption
{Two types of chaotic orbits (with constants ${\cal E}=0.956$, $\ell=4M$) found in the field of a black hole surrounded by the inverted 1st Morgan-Morgan disc with ${\cal M}=1.3M$, $r_{\rm disc}=20M$. Eight orbits filling the ``chaotic sea" on Poincar\'e section (top) have their $z$-position power spectra at bottom left, while four orbits dwelling mostly in the vicinity of regular islands in the centre have their spectra at bottom right.}
\label{spectra-4}
\end{figure}

``Chaoticity" of the system may surface in many respects. Its various characteristics tend to evolve in a more ``random" way when they are not bound by a sufficient number of isolating integrals. One of simple tests is thus to plot the time series of selected quantities and their Fourier spectra. For ``chaotic" systems, one expects the series to exhibit more randomness and the spectra to contain more ``lower harmonics" (possibly even blurred clusters of frequencies). We have picked the radial velocity and vertical position, and also the particle's ``latitudinal action", obtained by averaging the latitudinal four-velocity over an orbital period. (For these purposes, we followed the geodesics for a longer span, up to a coordinate time of about $10^6 M$ which typically corresponds to several thousands orbital periods.)

The easiest time series are those of phase variables. We have plotted the vertical position $z$ and the radial velocity $u^r$ plus their Fourier spectra for several orbits encountered in the previous ({\em Poincar\'e}) section. (We show only the $z$ behaviour here, but the $u^r$ series behave similarly.)
In Fig. \ref{spectra-1}, we pick two orbits around a black hole with the inverted 1st Morgan-Morgan disc and plot their spatial traces, Poincar\'e sections, time series of their $z$ position and the latter's Fourier spectra. For the regular orbit (which produces smooth curves on Poincar\'e surfaces), the time series follows regular oscillations, its Fourier spectrum only contains several distinctive peaks and goes over to nearly constant function at low frequencies. Chaotic orbit (which produces densely dotted, ``stochastic" layers on Poincar\'e surfaces), on the other hand, corresponds to complicated, inharmonic time series and fuzzy Fourier spectrum spanning over ``all" frequencies including the low ones. The irregularity of evolution can be seen at various scales, it need not be only ``local" --- from time to time, the orbital parameters may ``suddenly" change substantially and then calm down again for a longer period. This type of behaviour is very important for the evolution of astrophysical systems (it is known from the backtracking of the Solar system history, for example).

\cite{KoyamaKK-07} found, in studying the dynamics of spin particles in the Schwarzschild field, that two cases can be further distinguished within irregular orbits according to the low-frequency shape of their time-series spectra. The ``highly chaotic" orbits (filling large area --- the ``chaotic sea" --- in Poincar\'e sections) show ``white noise" there, i.e. the noise with relatively low and approximately constant mean value. On the other hand, the ``less chaotic" orbits (filling just narrow chaotic layers, namely confined to the vicinity of regular islands --- this case is being called ``sticky motion") show the $1/f$ dependence at low frequencies (the spectrum begins at higher amplitudes and goes down with frequency). We have checked whether similar behaviours can be found in our systems, too.

In figure \ref{spectra-2}, we compare several rather low-energy (${\cal E}=0.953$) orbits from the ``moderately chaotic" phase space of a black hole surrounded by the inverted 1st Morgan-Morgan disc (${\cal M}=0.5M$, $r_{\rm disc}=20M$). First we took ``most regular" orbits that live deep in the primary island, almost unaffected by perturbation (pink, light blue and violet). Their spectrum is very smooth, containing just a few peaks at nearly the same frequencies. The second triple (orange, blue and green) comes from the regular island of periodicity $k=3$. The spectra are again smooth with distinct peaks at similar frequencies, now containing more of them (these orbits only exist due to the perturbation and are more complicated then those belonging to the primary island). ``Even less regular" are the trajectories living on the outskirts of the primary island, encircling also the higher-periodic (e.g. the $k=3$) orbits and neighbouring with the chaotic layer. (These trajectories break up into the chaotic sea when energy is increased.) Their spectra already contain much more harmonics (near the separatrices, the resonances are very close to each other), but the low-frequency part is still smooth and constant. Finally, we took three orbits from the stochastic layer which, however, keep close to the $k=2$ islands and so might be only ``weakly chaotic" (black, red and green on the second Poincar\'e section on the right; mainly the black trajectory is seen to linger in the vicinity of the regular islands for a long time). The spectra of all these orbits are clearly chaotic and show the $1/f$ dependence at low frequencies.

The figures confirm that the results of \cite{KoyamaKK-07} apply to our systems as well. We wanted to see in further detail the difference between the ``almost regular" black trajectory, sticking to the $k=2$ island for a considerable amount of time, and the truly regular trajectory belonging to that island. In figure \ref{spectra-3}, one can see clear difference between time series of $z$ obtained for these seemingly very similar orbits. However, in proper times between $\tau=6.7\cdot 10^5M$ and $\tau=9.4\cdot 10^5M$ the chaotic orbit really behaves very much like a regular one. This is also confirmed by spatial picture of the orbits in the figure: in the above stage, the chaotic orbit is almost indistinguishable from the regular one, although it clearly looks chaotic in other times.

\cite{KoyamaKK-07} suggested that the degree of orbital chaoticity depends on strength of the perturbation (magnitude of the particles' spin in their case) rather than on energy. In order to check this on our system, we compared a few more orbits computed in the field with much heavier disc (${\cal M}=1.3M$) and with constants of motion for which the phase space is strongly chaotic. In figure \ref{spectra-4}, eight orbits filling the chaotic sea are compared with four orbits spending most of the time in the vicinity of regular islands. Power spectra of the $z$-position time series are different and indicate that the first set belongs to the white-noise type, whereas the second set belongs to the $1/f$ type. Hence, even in a more strongly perturbed and thus rather chaotic system there exist trajectories that ``mimic" the behaviour of regular ones for a certain amount of time (the action changes only slightly there), and this reflects in the $1/f$ shape of their low-frequency power spectra, even if they may inhabit larger portions of phase space at other times. The power-spectrum behaviour is thus a property of the phase-space layer to which the given orbit belongs, rather than of the space-time --- or the dynamical system --- as such.

Let us add that we obtained similar results for the other discs (other than the inverted 1st Morgan-Morgan), too.

\section{``Latitudinal action" integral}

\begin{figure}
\includegraphics[width=\textwidth]{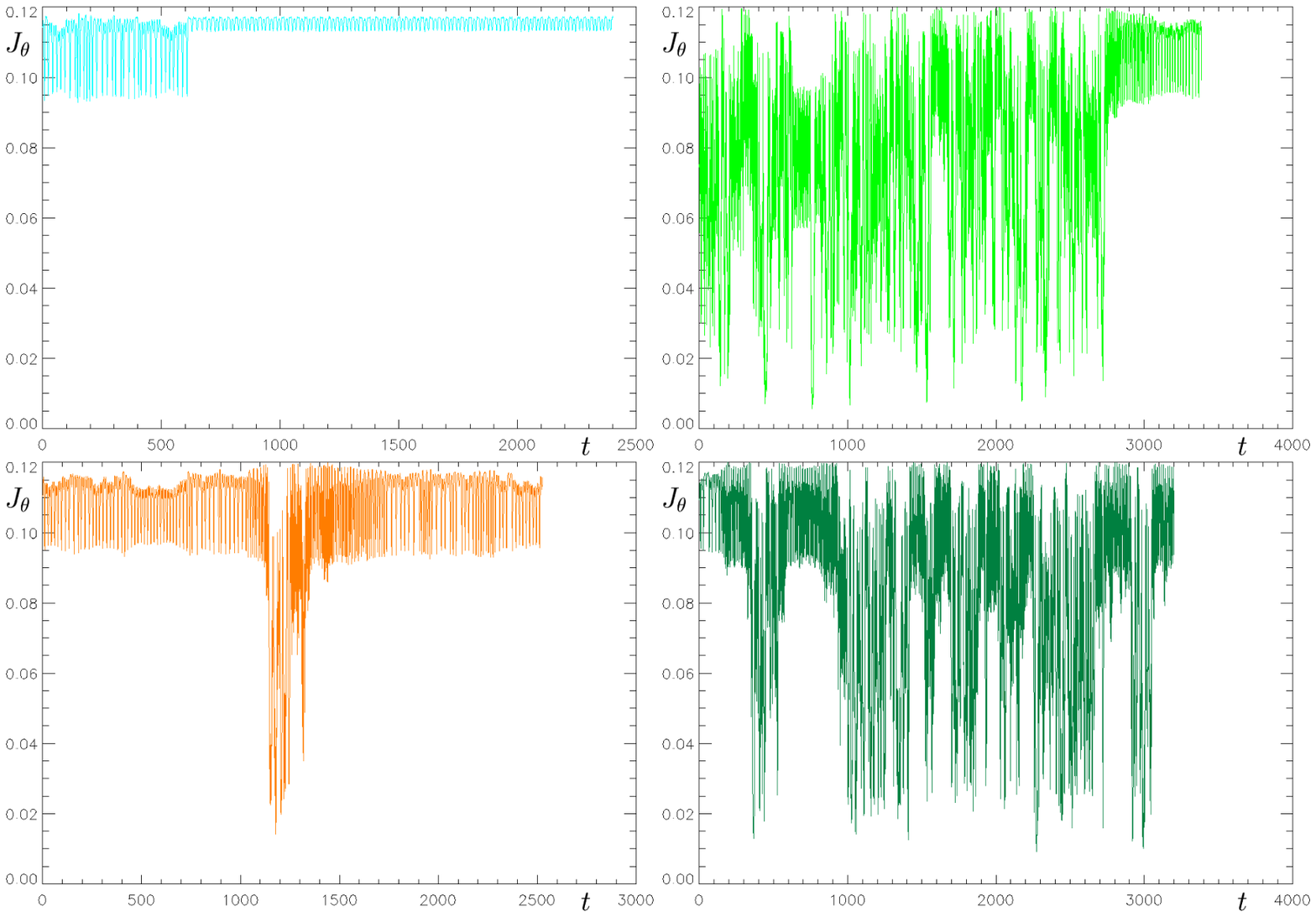}
\caption
{Time series of the latitudinal action (\ref{Jtheta}) as registered for the four weakly chaotic trajectories from fig. \ref{spectra-4}. (Fourier spectra of their $z$-position times series are given at bottom right of that figure.) The colours are kept the same as there. To be specific, we plot the action integral as divided by the length of the given orbital loop in terms of proper-time lapse.}
\label{lat-action}
\end{figure}

\begin{figure}
\includegraphics[width=\textwidth]{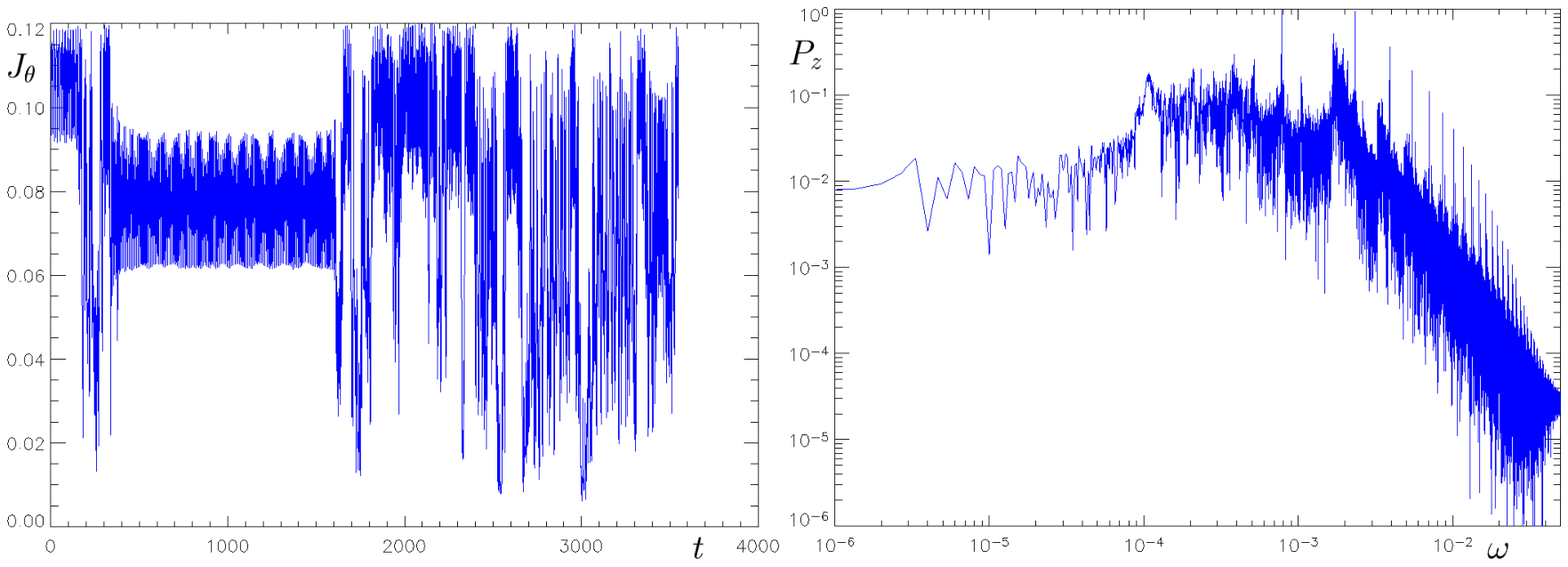}
\caption
{Time series of the latitudinal action (left) and $z$-position power spectrum (right) for the ``dark blue" orbit from fig. \ref{spectra-4}. The orbit spends quite a long time very near the regular islands of periodicity $k=3$ (they are not far from the boundary of the accessible region), but its power spectrum still does not show any signs of the 1/frequency behaviour at low frequencies.}
\label{lat-action-peculiar}
\end{figure}

There are various other variables whose time series should reflect the nature of the system dynamics. For example, one could register the time intervals between particle's returns to a given $r$ or a given $\theta$, the times between reaching turning points, etc.
In a paper on charged-particle motion in an EM field surrounding a rotating black hole, \cite{TakahashiK-09} considered the ``latitudinal action"
\begin{equation}  \label{Jtheta}
  J_\theta\equiv
  \frac{1}{l}\oint\sqrt{g_{\theta\theta}u^\theta u^\theta}\,{\rm d}\tau \;,
\end{equation}
given by averaging the test-particle latitudinal four-velocity over one orbital cycle (of length $l$). They compared its time evolution with Poincar\'e maps and evolution of $\theta$ and really observed clear correlation. We were curious whether a similar observation can be made in our disc/ring-perturbed systems, too.

For regular orbits, we obtained regular oscillations of the action around certain constant mean value. The amplitude of these oscillations is very small for orbits close to the primary resonance (``circular" periodic orbit) and larger for orbits belonging to higher-periodicity islands. For chaotic orbits, the action integral does not evolve in a harmonic way and is not tied to any obvious mean value.
In figure \ref{lat-action}, the behaviour of action is drawn for the four ``weakly chaotic" orbits from figure \ref{spectra-4} (spectra of their $z$-position are given at bottom right of that figure). All these trajectories are ``stuck on" regular islands in certain stages of their evolution --- and it is seen now that in those stages their action really behaves almost like along regular orbits (almost regular moderate oscillations around almost constant mean value).

Let us conclude with a trajectory that also adheres to a regular island for a certain time, but its power spectrum still does {\em not} show the $1/f$ shape at low frequencies. In figure \ref{spectra-4}, it is the dark blue geodesic occurring in the vicinity of the $k=3$ islands. In figure \ref{lat-action-peculiar}, we plot the time series of its latitudinal action and spectrum of its $z$-position evolution. Clearly the action behaves quite regularly for a considerable time (mainly in periods from $0$ to $150M$ and from $400M$ to $1600M$), yet the $z$-motion spectrum is still of clear ``white-noise" type, not showing any signs of the $1/f$ dependence. Perhaps the latter would only appear if the spectrum were calculated solely for the ``almost regular" phase of the orbit.

\section{Concluding remarks}

We have tried to get a basic idea about how the dynamics of free massive test particles in a Schwarzschild black-hole field is perturbed by gravity of a static, axially symmetric and reflectionally symmetric ring or disc. Poincar\'e sections have proved that it becomes chaotic and that the degree of stochasticity grows with compactness and relative mass of the external source and with energy of the particles. However, for large values of these parameters, the dynamics rather returns towards regular regime. The shape of stochastic regions depend very much on radius of the additional source. This is clearly the consequence of how the parameters affect the shape of effective potential, in particular of whether and where are regions that allow ``permanent" orbital motion. Our system shows typical features of a nearly-integrable Hamiltonian system, the behaviour revealed on Poincar\'e sections is a good illustration of the KAM theory. (Strictly speaking, the latter should only be applied to orbits which do not cross the disc, however.)

The appearance of selected orbits in Poincar\'e sections were compared with their spatial picture, with the time series representing evolution of their vertical ($z$) position, with the latter's Fourier spectra and with the evolution of the ``latitudinal action" assigned to their individual loops (fixed by $\theta$-periodicity). Chaotic orbits are clearly distinguishable from regular in all these respects. Within the chaotic trajectories, one can further recognise those which adhere to regular orbits, lingering within narrow bands in Poincar\'e diagrams for most of the time, and those which fill a large area (``chaotic sea") there in a rather uniform way. The degree of chaoticity is reflected in time series and their Fourier spectra: the ``weakly chaotic" orbits produce nearly regular and moderate oscillations around nearly constant mean value, which typically gives rise to ``1/frequency" shape of the spectrum at low frequencies, whereas ``strongly chaotic" orbits produce strongly irregular times series without any distinct mean value, which corresponds to ``white-noise" shape at low end of the spectrum (relatively flat curve at rather low values). The most interesting are chaotic orbits in strongly perturbed systems which undergo both the ``sticky motion" near regular islands and ``unrestrained" phases of drifting around the chaotic sea. The power spectra of their evolution usually ``take notice" of the almost regular phase and fall under the $1/f$ type, but we also came across a trajectory whose spectrum showed a clear white noise, although it behaved quite orderly for a considerable time.

The study of any specific dynamical system may develop in several obvious ways. First, there should be a running check of genericity/stability of the dynamical regime found. This involves going through the initial conditions in an exhaustive manner, changing the parameters of the system, trying different numerical schemes\dots Second, one may elaborate the model of the explored system in order to make it as realistic as possible, try to estimate and examine the consequences and mutual couplings of different ``perturbations"/neglects as well as their astrophysical importance. Finally, it is very appropriate to study the dynamics of the specific system by more methods and observe the relations between different aspects of its chaoticity. This is our next plan.

Concerning the last point(s), let us note that there especially arises the question about the role of features which are specific for the theory that underlies the phase space. In general relativity, such a special feature is the space-time curvature. It is clear that chaos is {\em not} solely caused by space-time curvature. (See the quotation from \cite{VieiraL-96b} in Sect. \ref{perturbing-background}. As a matter of fact, some systems are chaotic both in the Newton's theory, in special relativity and in general relativity.) However, the example of the Einstein's theory of gravitation itself suggests that it might be possible to interpret the dynamics of some systems as geodesic flows in suitably defined manifolds. This would certainly depend on the type of interaction, but also on some integral properties of the specific trajectory (like constants of the motion). This direction has been notably pursued by Szyd{\l}owski and coworkers, see e.g. \cite{Szydlowski-98,Szydlowski-99} and references therein, while dynamics of specific systems were interpreted ``geometrically" in some of the papers already cited in Sect. \ref{perturbing-background} \citep{Yurtsever-95,SotaSM-96} (cf. section 2.3 of \cite{CornishG-97}).
More recent contributions to this line of research include mainly \cite{CasettiPC-00,RamasubramanianS-01,ClementiP-02,Kawabe-05,HorwitzZLSL-07,
Cafaro-08,ZionH-08,CerrutiSolaCFP-08} (further references are given therein).

\section*{Acknowledgements}

We are grateful to M. \v{Z}\'a\v{c}ek who kindly provided us his numerical code, and to L. \v{S}ubr and D. Heyrovsk\'y for hints and discussions. O.S. thanks professor F. de Felice for hospitality at the Department of Physics, University of Padova.
The work was supported by the grants GACR-202/09/0772 (O.S.) and GAUK-86508 (P.S.) and by the Czech Ministry of Education under the projects MSM0021610860 and LC06014.


\begin{thebibliography}{99}

\bibitem[\protect\citeauthoryear{Aguirregabiria}{1997}]
        {Aguirregabiria-97}
   Aguirregabiria J. M., 1997,
   Chaotic scattering around black holes,
   Phys. Lett. A, 224, 234
\bibitem[\protect\citeauthoryear{Alonso et al.}{2008}]
        {AlonsoRS-08}
   Alonso D., Ruiz A., S\'anchez-Hern\'andez M., 2008,
   Escape of photons from two fixed extreme Reissner-Nordstr\"om black holes,
   Phys. Rev. D, 78, 104024
\bibitem[\protect\citeauthoryear{Andriopoulos \& Leach}{2008}]
        {AndriopoulosL-08}
   Andriopoulos K., Leach P. G. L., 2008,
   The Mixmaster Universe: the final reckoning?,
   J. Phys. A, 41, 155201
\bibitem[\protect\citeauthoryear{Bach \& Weyl}{1922}]
        {BachW-22}
   Bach R., Weyl H., 1922,
   Neue L\"osungen der Einsteinschen Gravitationsgleichungen.
   B. Explizite Aufstellung statischer axialsymmetrischer Felder,
   Math. Zeit., 13, 134
\bibitem[\protect\citeauthoryear{Benini \& Montani}{2004}]
        {BeniniM-04}
   Benini R., Montani G., 2004,
   Frame independence of the inhomogeneous mixmaster chaos
   via Misner-Chitr\'e-like variables,
   Phys. Rev. D, 70, 103527
\bibitem[\protect\citeauthoryear{Bombelli \& Calzetta}{1992}]
        {BombelliC-92}
   Bombelli L., Calzetta E., 1992,
   Chaos around a black hole,
   Class. Quantum Grav., 9, 2573
\bibitem[\protect\citeauthoryear{Buzzi et al.}{2007}]
        {BuzziLd-07}
   Buzzi C. A., Llibre J., da Silva P. R., 2007,
   On the dynamics of the Bianchi IX system,
   J. Phys. A, 40, 7187
\bibitem[\protect\citeauthoryear{Cafaro}{2008}]
        {Cafaro-08}
   Cafaro C., 2008,
   Information-geometric indicators of chaos in Gaussian models
   on statistical manifolds of negative Ricci curvature,
   Int. J. Theor. Phys., 47, 2924
\bibitem[\protect\citeauthoryear{Cardoso et al.}{2009}]
        {Cardoso-etal-09}
   Cardoso V., Miranda A. S., Berti E., Witek H., Zanchin V. T.,
   Geodesic stability, Lyapunov exponents and quasinormal modes,
   Phys. Rev. D, 79, 064016
\bibitem[\protect\citeauthoryear{Casetti et al.}{2000}]
        {CasettiPC-00}
   Casetti L., Pettini M., Cohen E. G. D., 2000,
   Geometric approach to Hamiltonian dynamics and statistical mechanics,
   Phys. Rep., 337, 237
\bibitem[\protect\citeauthoryear{Cerruti-Sola et al.}{2008}]
        {CerrutiSolaCFP-08}
   Cerruti-Sola M., Ciraolo G., Franzosi R., Pettini M., 2008,
   Riemannian geometry of Hamiltonian chaos: Hints for a general theory,
   Phys. Rev. E, 78, 046205
\bibitem[\protect\citeauthoryear{Chen \& Wang}{2003}]
        {ChenW-03}
   Chen J., Wang Y., 2003,
   Chaotic dynamics of a test particle around a gravitational field
   with a dipole,
   Class. Quantum Grav., 20, 3897
\bibitem[\protect\citeauthoryear{Clementi \& Pettini}{2002}]
        {ClementiP-02}
   Clementi C., Pettini M., 2002,
   A geometric interpretation of integrable motions,
   Celest. Mech. Dyn. Astron., 84, 263
\bibitem[\protect\citeauthoryear{\dots; Coley}{2002}]
        {Coley-02}
   Coley A. A., 2002,
   No chaos in brane-world cosmology,
   Class. Quantum Grav., 19, L45
\bibitem[\protect\citeauthoryear{Contopoulos \& Harsoula}{2005}]
        {ContopoulosH-05}
   Contopoulos G., Harsoula M., 2005,
   Chaotic motions in the field of two fixed black holes,
   Celest. Mech. Dyn. Astron., 92, 189
\bibitem[\protect\citeauthoryear{Contopoulos \& Papadaki}{1993}]
        {ContopoulosP-93}
   Contopoulos G., Papadaki H., 1993,
   Newtonian and relativistic periodic orbits around two fixed black holes,
   Celest. Mech. Dyn. Astron., 55, 47
\bibitem[\protect\citeauthoryear{Cornish \& Frankel}{1997}]
        {CornishF-97}
   Cornish N. J., Frankel N. E., 1997,
   The black hole and the pea,
   Phys. Rev. D, 56, 1903
\bibitem[\protect\citeauthoryear{Cornish \& Gibbons}{1997}]
        {CornishG-97}
   Cornish N. J., Gibbons G. W., 1997,
   A tale of two centres,
   Class. Quantum Grav., 14, 1865
\bibitem[\protect\citeauthoryear{Cornish \& Levin}{2003}]
        {CornishL-03}
   Cornish N. J., Levin J., 2003,
   Lyapunov timescales and black hole binaries,
   Class. Quantum Grav., 20, 1649
\bibitem[\protect\citeauthoryear{D'Afonseca et al.}{2005}]
        {D'AfonsecaLO-05}
   D'Afonseca L. A., Letelier P. S., Oliveira S. R., 2005,
   Geodesics around Weyl-Bach's ring solution,
   Class. Quantum Grav., 22, 3803
\bibitem[\protect\citeauthoryear{de Moura \& Letelier}{2000a}]
        {deMouraL-00a}
   de Moura A. P. S., Letelier P. S., 2000a,
   Chaos and fractals in geodesic motions around a nonrotating black hole
   with halos,
   Phys. Rev. E, 61, 6506
\bibitem[\protect\citeauthoryear{de Moura \& Letelier}{2000b}]
        {deMouraL-00b}
   de Moura, A. P. S., Letelier P. S., 2000b,
   Scattering map for two black holes,
   Phys. Rev. E, 62, 4784
\bibitem[\protect\citeauthoryear{Dettmann et al.}{1994}]
        {DettmannFC-94}
   Dettmann C. P., Frankel N. E., Cornish N. J., 1994,
   Fractal basins and chaotic trajectories in multi-black-hole spacetimes,
   Phys. Rev. D, 50, R618
\bibitem[\protect\citeauthoryear{Dettmann et al.}{1995}]
        {DettmannFC-95}
   Dettmann C. P., Frankel N. E., Cornish N. J., 1995,
   Chaos and fractals around black holes,
   Fractals, 3, 161
\bibitem[\protect\citeauthoryear{Dubeibe et al.}{2007}]
        {DubeibePS-07}
   Dubeibe F. L., Pach\'on L. A., Sanabria-G\'omez J. D., 2007,
   Chaotic dynamics around astrophysical objects with nonisotropic stresses,
   Phys. Rev. E., 75, 023008
\bibitem[\protect\citeauthoryear{Faraoni et al.}{2006}]
        {FaraoniJT-06}
   Faraoni V., Jensen M. N., Theuerkauf S. A., 2006,
   Non-chaotic dynamics in general-relativistic and scalar tensor cosmology,
   Class. Quantum Grav., 23, 4215
\bibitem[\protect\citeauthoryear{Fay \& Lehner}{2004}]
        {FayL-04}
   Fay S., Lehner T., 2004,
   Bianchi type IX asymptotical behaviours with a massive scalar field:
   chaos strikes back,
   Gen. Rel. Grav., 36, 2635
\bibitem[\protect\citeauthoryear{Grossman \& Levin}{2009}]
        {GrossmanL-09}
   Grossman R., Levin J., 2009,
   Dynamics of black hole pairs.
   II. Spherical orbits and the homoclinic limit of zoom-whirliness,
   Phys. Rev. D, 79, 043017
\bibitem[\protect\citeauthoryear{Gu\'eron \& Letelier}{2001}]
        {GueronL-01}
   Gu\'eron E., Letelier P. S., 2001,
   Chaos in pseudo-Newtonian black holes with halos,
   Astron. Astrophys., 368, 716
\bibitem[\protect\citeauthoryear{Gu\'eron \& Letelier}{2002}]
        {GueronL-02}
   Gu\'eron E., Letelier P. S., 2002,
   Geodesic chaos around quadrupolar deformed centers of attraction,
   Phys. Rev. E, 66, 046611
\bibitem[\protect\citeauthoryear{Han}{2008a}]
        {Han-08a}
   Han W., 2008a,
   Chaos and dynamics of spinning particles in Kerr spacetime,
   Gen. Rel. Grav., 40, 1831
\bibitem[\protect\citeauthoryear{Han}{2008b}]
        {Han-08b}
   Han W., 2008b,
   Revised research about chaotic dynamics in Manko et al. spacetime,
   Phys. Rev. D, 77, 123007
\bibitem[\protect\citeauthoryear{Hanan \& Radu}{2007}]
        {HananR-07}
   Hanan W., Radu E., 2007,
   Chaotic motion in multi-black hole spacetimes and holographic screens,
   Mod. Phys. Lett. A, 22, 399
\bibitem[\protect\citeauthoryear{Hartl}{2003a}]
        {Hartl-03a}
   Hartl M. D., 2003a,
   Dynamics of spinning test particles in Kerr spacetime,
   Phys. Rev. D, 67, 024005
\bibitem[\protect\citeauthoryear{Hartl}{2003b}]
        {Hartl-03b}
   Hartl M. D., 2003b,
   Survey of spinning test particle orbits in Kerr spacetime,
   Phys. Rev. D, 67, 104023
\bibitem[\protect\citeauthoryear{Heinzle et al.}{2006}]
        {HeinzleRU-06}
   Heinzle J. M., R\"ohr N., Uggla C., 2006,
   Homoclinic chaos and energy condition violation,
   Phys. Rev. D, 74, 061502
\bibitem[\protect\citeauthoryear{Heinzle \& Uggla}{2009}]
        {HeinzleU-09}
   Heinzle, J. M., Uggla C., 2009,
   Mixmaster: fact and belief,
   Class. Quantum Grav., 26, 075016
\bibitem[\protect\citeauthoryear{Hobill et al.}{1994}]
        {HobillBC-94}
   Hobill D., Burd A., Coley A., eds, 1994,
   Deterministic Chaos in General Relativity
   (NATO ASI Series 332).
   Plenum Press, New York and London
\bibitem[\protect\citeauthoryear{Horwitz et al.}{2007}]
        {HorwitzZLSL-07}
   Horwitz L., Zion Y. B., Lewkowicz M., Schiffer M., Levitan J., 2007,
   Geometry of Hamiltonian chaos,
   Phys. Rev. Lett., 98, 234301
\bibitem[\protect\citeauthoryear{Howard \& Wilkerson}{1995}]
        {HowardW-95}
   Howard J. E., Wilkerson T. D., 1995,
   Problem of two fixed centers and a finite dipole,
   Phys. Rev. A, 52, 4471
\bibitem[\protect\citeauthoryear{Hrycyna \& Szyd{\l}owski}{2006}]
        {HrycynaS-06}
   Hrycyna O., Szyd{\l}owski M., 2006,
   Different faces of chaos in FRW models with scalar fields --
   geometrical point of view,
   Chaos, Solitons \& Fractals, 28, 1252
\bibitem[\protect\citeauthoryear{Jor\'as \& Stuchi}{2003}]
        {JorasS-03}
   Jor\'as S. E., Stuchi T. J., 2003,
   Chaos in a closed Friedmann-Robertson-Walker universe:
   An imaginary approach,
   Phys. Rev. D, 68, 123525
\bibitem[\protect\citeauthoryear{Judd}{2008}]
        {Judd-08}
   Judd K., 2008,
   Shadowing pseudo-orbits and gradient descent noise reduction,
   J. Nonlinear Sci., 18, 57
\bibitem[\protect\citeauthoryear{Kao \& Cho}{2005}]
        {KaoCh-05}
   Kao J.-K., Cho H. T., 2005,
   The onset of chaotic motion of a spinning particle
   around the Schwarzchild  black hole,
   Phys. Lett. A, 336, 159
\bibitem[\protect\citeauthoryear{Karas \& \v{S}ubr}{2007}]
        {KarasS-07}
   Karas V., \v{S}ubr L., 2007,
   Enhanced activity of massive black holes by stellar capture assisted by
   a self-gravitating accretion disc,
   Astron. Astrophys., 470, 11
\bibitem[\protect\citeauthoryear{Karas \& Vokrouhlick\'y}{1992}]
        {KarasV-92}
   Karas V., Vokrouhlick\'y D., 1992,
   Chaotic motion of test particles in the Ernst space-time,
   Gen. Rel. Grav., 24, 729
\bibitem[\protect\citeauthoryear{Kawabe}{2005}]
        {Kawabe-05}
   Kawabe T., 2005,
   Indicator of chaos based on the Riemannian geometric approach,
   Phys. Rev. E, 71, 017201
\bibitem[\protect\citeauthoryear{Kiuchi \& Maeda}{2004}]
        {KiuchiM-04}
   Kiuchi K., Maeda K., 2004,
   Gravitational waves from a chaotic dynamical system,
   Phys. Rev. D., 70, 064036
\bibitem[\protect\citeauthoryear{Kiuchi et al.}{2007}]
        {KiuchiKM-07}
   Kiuchi K., Koyama H., Maeda K., 2007,
   Gravitational wave signals from a chaotic system: A point mass with a disk,
   Phys. Rev. D, 76, 024018
\bibitem[\protect\citeauthoryear{Koyama et al.}{2007}]
        {KoyamaKK-07}
   Koyama H., Kiuchi K., Konishi T., 2007,
   $1/f$ fluctuations in spinning-particle motions around
   a Schwarzschild black hole,
   Phys. Rev. D, 76, 064031
\bibitem[\protect\citeauthoryear{Lai et al.}{1999}]
        {LaiGK-99}
   Lai Y.-C., Grebogi C., Kurths J., 1999,
   Modeling of deterministic chaotic systems,
   Phys. Rev. E, 59, 2907
\bibitem[\protect\citeauthoryear{Lemos \& Letelier}{1994}]
        {LemosL-94}
   Lemos J. P. S., Letelier P. S., 1994,
   Exact general relativistic thin disks around black holes,
   Phys. Rev. D, 49, 5135
\bibitem[\protect\citeauthoryear{Letelier \& Vieira}{1997a}]
        {LetelierV-97a}
   Letelier P. S., Vieira W. M., 1997a,
   Chaos and rotating black holes with halos,
   Phys. Rev. D, 56, 8095
\bibitem[\protect\citeauthoryear{Letelier \& Vieira}{1997b}]
        {LetelierV-97b}
   Letelier P. S., Vieira W. M., 1997b,
   Chaos in black holes surrounded by gravitational waves,
   Class. Quantum Grav., 14, 1249
\bibitem[\protect\citeauthoryear{Letelier \& Vieira}{1998}]
        {LetelierV-98}
   Letelier P. S., Vieira W. M., 1998,
   Chaos and Taub-NUT related spacetimes,
   Phys. Lett. A, 244, 324
\bibitem[\protect\citeauthoryear{Levin}{1999}]
        {Levin-99}
   Levin J., 1999,
   Chaos may make black holes bright,
   Phys. Rev. D, 60, 064015
\bibitem[\protect\citeauthoryear{Levin}{2000}]
   {Levin-00}
   Levin J., 2000,
   Gravity waves, chaos, and spinning compact binaries,
   Phys. Rev. Lett., 84, 3515
\bibitem[\protect\citeauthoryear{Levin}{2003}]
        {Levin-03}
   Levin J., 2003,
   Fate of chaotic binaries,
   Phys. Rev. D, 67, 044013
\bibitem[\protect\citeauthoryear{Levin}{2006}]
        {Levin-06}
   Levin J., 2006,
   Chaos and order in models of black hole pairs,
   Phys. Rev. D, 74, 124027
\bibitem[\protect\citeauthoryear{Levin \& Perez-Giz}{2009}]
        {LevinPG-09}
   Levin J., Perez-Giz G., 2009,
   Homoclinic orbits around spinning black holes.
   I. Exact solution for the Kerr separatrix,
   Phys. Rev. D, 79, 124013
\bibitem[\protect\citeauthoryear{Lichtenberg \& Lieberman}{1992}]
        {LichtenbergL-92}
   Lichtenberg A. J., Lieberman M. A., 1992,
   Regular and Chaotic Dynamics, 2nd edn.
   Springer, Berlin
\bibitem[\protect\citeauthoryear{Lukes-Gerakopoulos et al.}{2008}]
        {LukesGBC-08}
   Lukes-Gerakopoulos G., Basilakos S., Contopoulos G., 2008,
   Dynamics and chaos in the unified scalar field cosmology,
   Phys. Rev. D, 77, 043521
\bibitem[\protect\citeauthoryear{Maciejewski et al.}{2008}]
        {MaciejewskiPSS-08}
   Maciejewski A. J., Przybylska M., Stachowiak T., Szyd{\l}owski M., 2008,
   Global integrability of cosmological scalar fields,
   J. Phys. A, 41, 465101
\bibitem[\protect\citeauthoryear{Moeckel}{1992}]
        {Moeckel-92}
   Moeckel R., 1992,
   A nonintegrable model in general relativity,
   Commun. Math. Phys., 150, 415
\bibitem[\protect\citeauthoryear{Motter}{2003}]
        {Motter-03}
   Motter A. E., 2003,
   Relativistic chaos is coordinate invariant,
   Phys. Rev. Lett., 91, 231101
\bibitem[\protect\citeauthoryear{Motter \& Saa}{2009}]
        {MotterS-09}
   Motter A. E., Saa A., 2009,
   Relativistic invariance of Lyapunov exponents
   in bounded and unbounded systems,
   Phys. Rev. Lett., 102, 184101
\bibitem[\protect\citeauthoryear{Perez-Giz \& Levin}{2009}]
        {PerezGizL-09}
   Perez-Giz G., Levin J., 2009,
   Homoclinic orbits around spinning black holes.
   II. The phase space portrait,
   Phys. Rev. D, 79, 124014
\bibitem[\protect\citeauthoryear{Podolsk\'y \& Kofro\v{n}}{2007}]
        {PodolskyK-07}
   Podolsk\'y J., Kofro\v{n} D., 2007,
   Chaotic motion in Kundt spacetimes,
   Class. Quantum Grav., 24, 3413
\bibitem[\protect\citeauthoryear{Podolsk\'y \& Vesel\'y}{1998}]
        {PodolskyV-98}
   Podolsk\'y J., Vesel\'y K., 1998,
   Chaotic motion in {\it pp}-wave spacetimes,
   Class. Quantum Grav., 15, 3505
\bibitem[\protect\citeauthoryear{Ramasubramanian \& Sriram}{2001}]
        {RamasubramanianS-01}
   Ramasubramanian K., Sriram M. S., 2001,
   Global geometric indicator of chaos and Lyapunov exponents in
   Hamiltonian systems,
   Phys. Rev. E, 64, 046207
\bibitem[\protect\citeauthoryear{Ramos-Caro et al.}{2008}]
        {Ramos-CaroLG-08}
   Ramos-Caro J., L\'opez-Suspes F., Gonz\'alez G. A., 2008,
   Chaotic and regular motion around generalized Kalnajs discs,
   Mon. Not. R. Astron. Soc., 386, 440
\bibitem[\protect\citeauthoryear{Saa}{2000}]
        {Saa-00}
   Saa A., 2000,
   Chaos around the superposition of a monopole and a thick disk,
   Phys. Lett. A, 269, 204
\bibitem[\protect\citeauthoryear{Saa \& Venegeroles}{1999}]
        {SaaV-99}
   Saa A., Venegeroles R., 1999,
   Chaos around the superposition of a black-hole and a thin disk,
   Phys. Lett. A, 259, 201
\bibitem[\protect\citeauthoryear{Sauer et al.}{1997}]
        {SauerGY-97}
   Sauer T., Grebogi C., Yorke J. A., 1997,
   How long do numerical chaotic solutions remain valid?,
   Phys. Rev. Lett., 79, 59
\bibitem[\protect\citeauthoryear{Scott}{2007}]
        {Scott-07}
   Scott A. C., 2007,
   The Nonlinear Universe: Chaos, Emergence, Life.
   Springer, Berlin
\bibitem[\protect\citeauthoryear{\c{S}elaru et al.}{2005}]
        {SelaruMCG-05}
   \c{S}elaru D., Mioc V., Cucu-Dumitrescu C., Ghenescu M., 2005,
   Chaos in Hill's generalized problem: from the solar system to black holes,
   Astron. Nachr., 326, 356
\bibitem[\protect\citeauthoryear{Semer\'ak}{2003}]
        {Semerak-03}
   Semer\'ak O., 2003,
   Gravitating discs around a Schwarzschild black hole: III,
   Class. Quantum Grav., 20, 1613
\bibitem[\protect\citeauthoryear{Semer\'ak}{2004}]
        {Semerak-04}
   Semer\'ak O., 2004,
   Exact power-law discs around static black holes,
   Class. Quantum Grav., 21, 2203
\bibitem[\protect\citeauthoryear{Semer\'ak et al.}{1999}]
        {SemerakZZ-99}
   Semer\'ak O., \v{Z}\'a\v{c}ek M., Zellerin T., 1999,
   Test-particle motion in superposed Weyl fields,
   Mon. Not. R. Astron. Soc., 308, 705
\bibitem[\protect\citeauthoryear{Shi et al.}{2001}]
        {Shi-etal-01}
   Shi P., He D., Kang W., Fu W., Hu G., 2001,
   Chaoslike behavior in nonchaotic systems at finite computation precision,
   Phys. Rev. E, 63, 046310
\bibitem[\protect\citeauthoryear{Soares \& Stuchi}{2005}]
        {SoaresS-05}
   Soares I. D., Stuchi T. J., 2005,
   Homoclinic chaos in axisymmetric Bianchi IX cosmologies reexamined,
   Phys. Rev. D, 72, 083516
   (Erratum: 2006, 73, 069901)
\bibitem[\protect\citeauthoryear{Sota et al.}{1996}]
        {SotaSM-96}
   Sota Y., Suzuki S., Maeda K., 1996,
   Chaos in static axisymmetric spacetimes: I. Vacuum case,
   Class. Quantum Grav., 13, 1241
\bibitem[\protect\citeauthoryear{Steklain \& Letelier}{2006}]
        {SteklainL-06}
   Steklain A. F., Letelier P. S., 2006,
   Newtonian and pseudo-Newtonian Hill problem,
   Phys. Lett. A, 352, 398
\bibitem[\protect\citeauthoryear{Steklain \& Letelier}{2009}]
        {SteklainL-09}
   Steklain A. F., Letelier P. S., 2009,
   Stability of orbits around a spinning body
   in a pseudo-Newtonian Hill problem,
   Phys. Lett. A, 373, 188
\bibitem[\protect\citeauthoryear{Str\'ansk\'y et al.}{2006}]
        {StranskyKC-06}
   Str\'ansk\'y P., Kurian M., Cejnar P., 2006,
   Classical chaos in the geometric collective model,
   Phys. Rev. C, 74, 014306
\bibitem[\protect\citeauthoryear{Suzuki \& Maeda}{1997}]
        {SuzukiM-97}
   Suzuki S., Maeda K., 1997,
   Chaos in Schwarzschild spacetime: The motion of a spinning particle,
   Phys. Rev. D, 55, 4848
\bibitem[\protect\citeauthoryear{Suzuki \& Maeda}{1999}]
        {SuzukiM-99}
   Suzuki S., Maeda K., 1999,
   Signature of chaos in gravitational waves from a spinning particle,
   Phys. Rev. D, 61, 024005
\bibitem[\protect\citeauthoryear{Szyd{\l}owski}{1997}]
        {Szydlowski-97}
   Szyd{\l}owski M., 1997,
   On invariant qualitative chaos in multi-black-hole spacetimes,
   Int. J. Mod. Phys. D, 6, 741
\bibitem[\protect\citeauthoryear{Szyd{\l}owski}{1998}]
        {Szydlowski-98}
   Szyd{\l}owski M., 1998,
   The Eisenhart geometry as an alternative description of dynamics in terms
   of geodesics,
   Gen. Rel. Grav., 30, 887
\bibitem[\protect\citeauthoryear{Szyd{\l}owski}{1999}]
        {Szydlowski-99}
   Szyd{\l}owski M., 1999,
   Desingularization of Jacobi metrics and chaos in general relativity,
   J. Math. Phys., 40, 3519
\bibitem[\protect\citeauthoryear{Takahashi \& Koyama}{2009}]
        {TakahashiK-09}
   Takahashi M., Koyama H., 2009,
   Chaotic motion of charged particles in an electromagnetic field surrounding
   a rotating black hole,
   Astrophys. J., 693, 472
\bibitem[\protect\citeauthoryear{Varvoglis \& Papadopoulos}{1992}]
        {VarvoglisP-92}
   Varvoglis H., Papadopoulos D., 1992,
   Chaotic interaction of charged particles with a gravitational wave,
   Astron. Astrophys., 261, 664
\bibitem[\protect\citeauthoryear{Vieira \& Letelier}{1996a}]
        {VieiraL-96a}
   Vieira W. M., Letelier P. S., 1996a,
   Chaos around a H\'enon-Heiles-inspired exact perturbation of a black hole,
   Phys. Rev. Lett., 76, 1409
\bibitem[\protect\citeauthoryear{Vieira \& Letelier}{1996b}]
        {VieiraL-96b}
   Vieira W. M., Letelier P. S., 1996b,
   Curvature and chaos in general relativity,
   Class. Quantum Grav., 13, 3115
\bibitem[\protect\citeauthoryear{Vieira \& Letelier}{1999}]
        {VieiraL-99}
   Vieira W. M., Letelier P. S., 1999,
   Relativistic and Newtonian core-shell models:
   Analytical and numerical results,
   Astrophys. J., 513, 383
\bibitem[\protect\citeauthoryear{Vokrouhlick\'y \& Karas}{1998}]
        {VokrouhlickyK-98}
   Vokrouhlick\'y D., Karas V., 1998,
   Stellar dynamics in a galactic centre surrounded by a massive accretion
   disc --- I. Newtonian description,
   Mon. Not. R. Astron. Soc., 298, 53
\bibitem[\protect\citeauthoryear{Wu \& Huang}{2003}]
        {WuH-03}
   Wu X., Huang T.-Y., 2003,
   Computation of Lyapunov exponents in general relativity,
   Phys. Lett. A, 313, 77
\bibitem[\protect\citeauthoryear{Wu \& Xie}{2008}]
        {WuX-08}
   Wu X., Xie Y., 2008,
   Resurvey of order and chaos in spinning compact binaries,
   Phys. Rev. D, 77, 103012
\bibitem[\protect\citeauthoryear{Wu \& Zhang}{2006}]
        {WuZ-06}
   Wu X., Zhang H., 2006,
   Chaotic dynamics in superposed Weyl spacetime,
   Astrophys. J., 652, 1466
\bibitem[\protect\citeauthoryear{Yurtsever}{1995}]
        {Yurtsever-95}
   Yurtsever U., 1995,
   Geometry of chaos in the two-center problem in general relativity,
   Phys. Rev. D, 52, 3176
\bibitem[\protect\citeauthoryear{Zion \& Horwitz}{2008}]
        {ZionH-08}
   Zion Y. B., Horwitz L., 2008,
   Applications of geometrical criteria for transition to Hamiltonian chaos,
   Phys. Rev. E, 78, 036209

\end{thebibliography}
\end{document}